\documentclass[12pt]{article}

\usepackage{float}
\usepackage{graphics, graphicx}
\usepackage{amsfonts}
\usepackage{natbib}

\usepackage[margin=10pt,font=small,labelfont=bf,labelsep=period]{caption}

\usepackage{amsmath,amsbsy,amssymb,amsthm,latexsym, tabularx}
\usepackage{multirow,multicol,booktabs,url}

\newtheorem{thm}{Theorem}

\newtheorem{lemma}{Lemma}

\usepackage{enumitem}

\newcommand{\bt}{\begin{itemize}}
\newcommand{\et}{\end{itemize}}
\newcommand{\ben}{\begin{enumerate}}
\newcommand{\een}{\end{enumerate}}
\newcommand{\beq}{\begin{equation}}
\newcommand{\eeq}{\end{equation}}
\newcommand{\beqn}{\begin{eqnarray}}
\newcommand{\eeqn}{\end{eqnarray}}
\newcommand{\bea}{\begin{eqnarray*}}
\newcommand{\eea}{\end{eqnarray*}}

\usepackage{algorithm}
\usepackage[noend]{algpseudocode}

\usepackage{setspace}
\let\Algorithm\algorithm
\renewcommand\algorithm[1][]{\Algorithm[#1]\setstretch{0.96}}

\usepackage{graphicx}
\usepackage{float}
\usepackage{rotating}
\usepackage{caption} 
\usepackage{subcaption}  %(for showing 4 graphs on the same figure and putting (a), (b), etc underneath)

\usepackage{color,hyperref}
\definecolor{darkblue}{rgb}{0.0,0.0,0.55}
\hypersetup{colorlinks,breaklinks,
            linkcolor=darkblue,urlcolor=darkblue,
            anchorcolor=darkblue,citecolor=darkblue}
            
\definecolor{darkpastelgreen}{rgb}{0.01, 0.75, 0.24}

\renewenvironment{itemize}{
\begin{list}{}{
\setlength{\leftmargin}{4em}
}}{
  \end{list} }

\begin{document}

\title{
A composite likelihood ratio approach to the analysis of correlated binary data in genetic association studies
}

\author{
Zeynep Baskurt$^{1\ast}$,
Lisa Strug$^{1,2}$\\
\small{$^{1}$ Genetics and Genome Biology,
The Hospital for Sick Children
Toronto, Ont., Canada}\\
\small{$^{2}$Dalla Lana School of Public Health, University of Toronto, Toronto, Ont., Canada}\\
\footnotesize{$^{\ast}$ Corresponding Author: \href{mailto:zeynep.baskurt@sickkids.ca}{zeynep.baskurt@sickkids.ca}}
}

\date{}

\maketitle

\begin{abstract}
The likelihood function represents statistical evidence in the context of data and a probability model. Considerable theory has demonstrated that evidence strength for different parameter values can be interpreted from the ratio of likelihoods at different points on the likelihood curve. The likelihood function can, however, be unknown or difficult to compute; e.g. for genetic association studies with a binary outcome in large multi-generational families. Composite likelihood is a convenient alternative to using the real likelihood and here we show composite likelihoods have valid evidential interpretation. We show that composite likelihoods, with a robust adjustment, have two large sample performance properties that enable reliable evaluation of relative evidence for different values on the likelihood curve: (1) The composite likelihood function will support the true value over the false value by an arbitrarily large factor; and (2) the probability of favouring a false value over a true value with high probability is small and bounded. Using an extensive simulation study, and in a genetic association analysis of reading disability in large complex pedigrees, we show that the composite approach yields valid statistical inference. Results are compared to analyses using generalized estimating equations and show similar inference is obtained, although the composite approach results in a full likelihood solution that provides additional complementary information.
\end{abstract}


\newpage

\section{Introduction}\label{section:introduction}

Genetic association studies have identified many genes or markers that contribute to disease susceptibility  \citep{GWAS:catalog}. Genetic data from families are often collected for the purpose of linkage analysis, however this pedigree data can also be used for fine-mapping studies using population based association analysis \citep{browning:2005}. When related individuals are involved in the analysis, GEE or generalized linear mixed effects models are commonly implemented in a frequentist framework \citep{thornton:2015}. A comprehensive comparison of several methods, namely, GEE, generalized linear mixed model, and a variance component model, for genome-wide association studies was conducted in \citet{chen:2011}  and, after consideration for low disease prevalence and rare genetic variants, linear mixed models were the only approach that resulted in valid type I error and adequate statistical power in the majority of cases. However, this method does not have an odds ratio interpretation for the regression coefficients and is computationally challenging. GEE is computationally more efficient and the inflation in type I error for the GEE is due to small sample size that can be avoided by using a jackknife variance estimator. However, the GEE is an estimation equation approach and does not allow for full likelihood interpretation. In the next section, we review the likelihood paradigm \citep{royall:1997}, a paradigm for statistical inference directly from the likelihood (or pseudo-likelihood) function that provides an alternative approach to this problem.

\subsection{Likelihood Paradigm}\label{section:lik_paradigm}

The likelihood paradigm uses likelihood functions to represent the statistical evidence generated in a study about the unknown parameter(s) of interest and uses likelihood ratios to measure the strength of statistical evidence for one hypothesis versus another. Suppose we observe $y$ as a realization of a random variable $Y$ with a probability distribution $\{f(.;\theta),\theta \in \Theta\}$ where $\theta$ is a fixed dimensional parameter. The Law of Likelihood \citep{hacking:1965} states: `if hypothesis $H_1$ implies that the probability that a random variable $Y$ takes the value $y$ is $f_1(y)$, while hypothesis $H_2$ implies that the probability is $f_2(y)$, then the observation $Y=y$ is evidence supporting $H_1$ over $H_2$ if and only if $f_1(y) > f_2(y)$, and the likelihood ratio, $f_1(y)/f_2(y)$, measures the strength of that evidence'.
Then, for $L(\theta) \propto f(y;\theta)$, $L(\theta_1)/L(\theta_2)$ measures the strength of evidence in favour of $H_1:\theta=\theta_1$ relative to $H_2:\theta=\theta_2$.
 We have strong evidence in favour of $H_1$ versus $H_2$ if $L(\theta_1)/L(\theta_2)>k$, strong evidence in favor of $H_2$ versus $H_1$ if  $L(\theta_1)/L(\theta_2)<1/k$ and we have weak evidence if  $1/k <L(\theta_1)/L(\theta_2)< k$ that is, the data did not produce sufficiently strong evidence in favor of either hypothesis. This is an undesirable result since it tells us the data provided are not sufficient to produce strong evidence, we need to increase the sample size. Another undesirable result is to observe strong evidence in favour of the wrong hypothesis, that is, observing misleading evidence, which also can be minimized by increasing the sample size. Moreover, below we see that the probability of observing misleading evidence is bounded. The choice of $k$ can be determined in the planning stage such that the probability of observing weak and misleading evidence is small. For discussions on benchmarks for $k$, see \citet[p.11]{royall:1997}. 
 
\begin{figure} [H]
\centerline{ \rotatebox{0}{\resizebox{!}{10cm}{
  \includegraphics{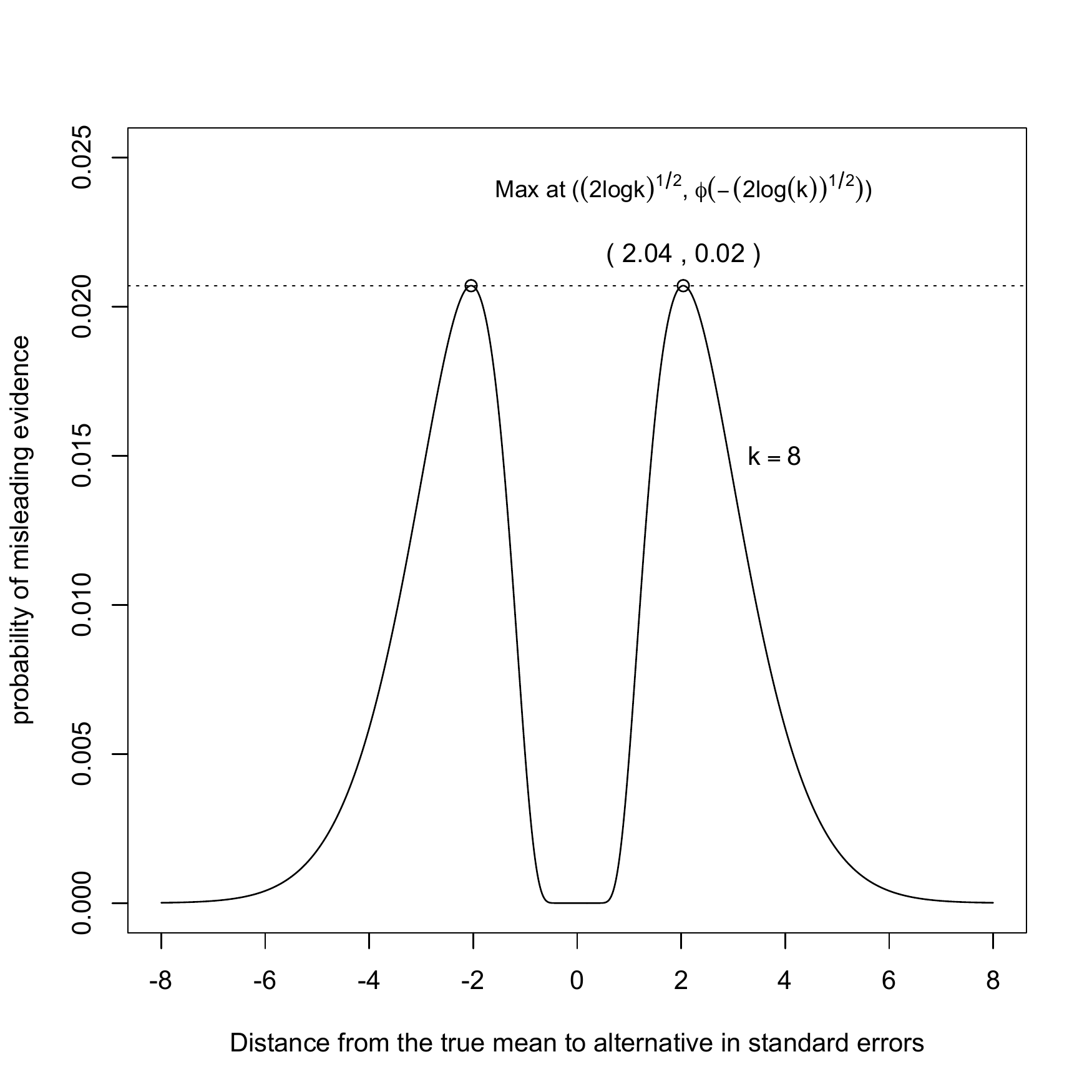}}}
   }
\caption{ Example of a bump function generated under the Normal distribution. \label{fig:bumpfunction}}
\end{figure}

The probability of getting misleading evidence, a function of $k$ and $n$, is denoted by $M_1(n,k)=P_1(L(\theta_2)/L(\theta_1) \geq k)$, where $P_1$ indicates the probability is taken under the correct model hypothesized in $H_1$. \citet{royall:2000} shows that this probability is described by a bump function, $P_1 (L(\theta_2)/L(\theta_1) \geq k) \to \Phi(-c/2-\log(k)/c)$, where $k>1$, $\Phi$ is the standard normal distribution function and $c$ is proportional to the distance between $\theta_1$ and $\theta_2$, where $c=\Delta \sqrt{n}/\sigma$. When the distance $\Delta$ is measured in units of the standard error, the probability of misleading evidence is independent of the sample size at a fixed $c$ (Figure \ref{fig:bumpfunction}).  Figure \ref{fig:bumpfunction} indicates that the probability of observing misleading evidence is 0 when the distance between the two hypothesized values is very small and corresponds to high probabilities of observing weak evidence. The probability of observing misleading evidence tends to 0 when the distance between the two hypothesized values increases and, regardless of the sample size, the bump function is maximized at $\Phi(-\sqrt{2\log k})$ when $\Delta=(2\log k)^{1/2}$ and this is the best possible bound.

To ensure reliable inferential properties for the likelihood paradigm, two important performance properties are required (\citet{royall:2000}, \citet{royall:2003}). Let $L(\theta)$ be the likelihood function where $X_1,...,X_n$ are iid with a smooth probability model $f(.;\theta)$ for $\theta \in \mathcal{R}$. Then 1) for any false value $\theta \neq \theta_0$, the evidence will eventually support $\theta_0$ over $\theta$ by an arbitrarily large factor: $P_0  ( L(\theta_0)/L(\theta) \to \infty \quad as \quad n \to \infty )=1$, and 2) in large samples, the probability of misleading evidence, as a function of $\theta$, is approximated by the bump function, $P_0 ( L(\theta)/L(\theta_0) \geq k) \to \Phi(-c/2-\log(k)/c)$ where $k>1$, $\Phi$ is the standard normal distribution function and $c$ is proportional to the distance between $\theta$ and $\theta_0$. The results can be extended to the case where $\theta$ is a fixed dimensional vector parameter.
The first property implies that the probability of getting strong
evidence in favor of the true value goes to 1. This implies that the probability of weak evidence and misleading evidence go to 0 as $n \to \infty$. The second property implies
that, when $n$ is sufficiently large, the probability of misleading evidence of
strength $k$ is maximized at a fixed constant $\Phi(-(2\log
k)^{1/2})$, over all $\theta$. Those properties ensure
that with high probability we will get evidence in favor of the true
value and that the probability of strong evidence in favor of a
false value is low.

In some situations, the working model for the data can be wrong.  \citet{royall:2003} show that under certain conditions, the likelihood ratio constructed from the working model, with a robust adjustment factor, can continue to be a valid measure for evidential interpretation. The first condition in order to use this likelihood ratio as a valid evidence function is to check whether \textit{the object of inference} is equal to \textit{the object of interest} for the chosen working model. We describe these concepts below.

Suppose we have $\{f(.;\theta),\theta \in \Theta\}$ where $\theta$ is fixed dimensional, as the working model, and there exist a true density  $g$. The Kullback-Liebler divergence between $g$ and $f$ is $K(g:f)= E_g \{\log g(.)- \log f(.:\theta)\} $. Let $\theta_g$ be the value of $\theta$ that maximizes $E_g\{\log f(.:\theta)\}$; that is, $\theta_g$ minimizes the Kullback-Liebler divergence between $f$ and $g$. Then it can be shown that the likelihood ratio $L(\theta_g)/L(\theta)$ constructed from $f$ converges to infinity with probability 1. This tells us that the likelihood under the wrong model represents evidence about $\theta_g$. Suppose we are interested in $E_g(Y)$. $\theta_g$ and $E_g(Y)$ are referred as \textit{the object of inference} and \textit{the object of interest} in \citet{royall:2003}. If our working model is wrong, then $\theta_g$ might not be equal to $E_g(Y)$, and the likelihood ratio will favour the wrong value $\theta_g$ over the true value $E_g(Y)$ since $P_g (L(\theta_g)/L(E_g(Y)  \to \infty)=1$. Thus, we essentially need to understand what $\theta_g$ represents in our working model $f$, and specifically, check if $\theta_g$ corresponds to $E_g(Y)$ once $f$ is chosen. This can be done analytically \citep{royall:2003} or through simulations (Section \ref{section:simulation_section}).

To allow for pure likelihood interpretation here we propose using composite likelihood ratios for genetic association studies when we have binary data with correlated outcomes. Our method provides full likelihood interpretation for the inference, i.e. support intervals, odds ratio (OR) interpretation, shape of the function and ability to compare the relative evidence for all parameter values. It is easy to implement and flexible to incorporate any data structure including independent controls.
 
\section{Methods}\label{section:methods}
\subsection{Composite Likelihoods : Definitions and Notations }
Composite likelihoods are constructed by multiplying together lower dimensional likelihood
objects \citep{lindsay:1988}. They are useful for inference when
the full likelihood is intractable or impractical to construct.
Suppose $\mathbf{Y}=(Y_1,Y_2,...,Y_m)$ is an $m-$dimensional random variable with  a specified joint density function, $f(\mathbf{y};\theta)$, where $\theta \in \Theta \subset \mathcal{R}^d$ is some unknown parameter. Considering this parametric model 
and a set of measurable events $\{\mathcal{A}_k; i=1,...,K\}$,  a composite likelihood is defined as $CL(\theta; \mathbf{y})=\prod_{k=1}^K f(\mathbf{y} \in  \mathcal{A}_k ;\theta)^{w_k}$, where $w_k, \; k=1,\ldots, K$ are positive weights.

The associated composite log-likelihood is denoted by $cl(\theta;\mathbf{y})=\log CL(\theta; \mathbf{y})$ following the notation in \cite{varin:2008}.  When we consider a random sample of size $n$,  the composite likelihood becomes $CL(\theta; \underset{\sim}{\mathbf{y}})  =  CL(\theta) = \prod_{i=1}^n CL(\theta; \mathbf{y}_i)  =  \prod_{i=1}^n \prod_{k=1}^K f(\mathbf{y}_i \in \mathcal{A}_k;\theta)^{w_k}$ with the \textit{composite score function} $u(\theta; \underset{\sim}{\mathbf{y}}) = u(\theta) = \Delta_\theta~cl(\theta)$, where $\Delta_\theta$ is the differentiation operation with respect to the parameter $\theta$. 
In the following, we drop the argument $\underset{\sim}{\mathbf{y}}$ for notational simplicity. 

Note that the parametric statistical model $\{ f(\mathbf{y};\theta), \; \mathbf{y} \in \mathcal{Y} \subset \mathcal{R}^m, \theta \in \Theta \subset \mathcal{R}^d  \}$ may or may not contain the true density $g(\mathbf{y})$ of $\mathbf{Y}$. \citet*{varin:2005} defined the \textit{Composite Kullback-Leibler divergence} between the assumed model $f$ and the true model $g$ as $K(g : f ; \theta)=\sum_{k=1}^K E_g \{ \log g(Y \in A_k) -\log f(Y \in A_k;\theta) \}w_k$. This is a linear combination of the Kullback-Leibler divergence associated with individual components of the composite likelihood. In the case where $f(y \in A_k;\theta)\neq g_k(y)$ for some $k$, the estimating equation $u(\theta)=0$ is not unbiased, i.e. $E_g\{u(\theta)\} \neq 0\quad \forall \theta$.  However, for the parameter value $\theta_g$, which uniquely minimizes the composite Kullback-Leibler divergence, $E_g\{u(\theta_g)\}= 0$ holds. Then, under some regularity conditions, the maximum composite likelihood estimator (MCLE), $\hat{\theta}_{CL}=\arg \max CL(\theta)$, converges to this pseudo-true value $\theta_g$. Note that $\theta_g$ depends on the choice of $A_k$. \citet*{xu:2012} provided a rigorous proof of the $\theta_g$-- consistency of $\hat{\theta}_{CL}$ under model misspecification. Furthermore, when $f(y \in A_k;\theta_0)= g_k(y)$ for all $k$, $\hat{\theta}_{CL}$ is a consistent estimator of the true parameter value $\theta_0$ \citep{xu:2012}.

In many practical settings, the parameter of interest is only a subset of the parameter space. In such cases, we partition $\theta$ into $\theta=(\psi,\lambda) \in \Theta \subset \mathcal{R}^d$, where $\psi \in \mathcal{R}^p$ is the parameter of interest and $\lambda \in \mathcal{R}^q$ is the nuisance parameter, with $p+q=d$. 
Then, $\hat{\theta}_{\psi}=(\psi,\hat{\lambda}(\psi))$ denotes the constrained MCLE of $\theta$ for fixed $\psi$, and $CL_p(\psi)$ is the profile composite likelihood function, $CL_p(\psi)=CL(\hat{\theta}_\psi)=\max_\lambda CL(\psi,\lambda)$.
The composite score function is partitioned as $u(\theta)=[ u_{\psi}(\theta)  \quad u_{\lambda}(\theta) ]^t=[ \partial cl (\theta)/\partial \psi   \quad  \partial cl (\theta)/\partial \lambda  ]^t$.
Taking the expectation of its second moments, we obtain the variability matrix,
$$J=\left[\begin{array}{cc}
J_{\psi \psi} & J_{\psi \lambda} \\
J_{\lambda \psi} & J_{\lambda \lambda} 
\end{array}
\right],$$
with $J_{\psi \psi}=E_g\{ (\partial cl(\theta;\mathbf{Y})/ \partial \psi)^2  \}$ and $J_{\psi \lambda}=E_g\{(\partial cl(\theta;\mathbf{Y})/ \partial \psi)(\partial cl(\theta)/ \partial \lambda) \}$. 
The information in the composite score function is given by $G(\theta)=H(\theta)J(\theta)^{-1}H(\theta)$ where $H$ is the sensitivity matrix defined along with its inverse. 
$$H=\left[\begin{array}{cc}
H_{\psi \psi} & H_{\psi \lambda} \\
H_{\lambda \psi} & H_{\lambda \lambda} 
\end{array}
\right],   \quad \quad \quad \quad H^{-1}=\left[\begin{array}{cc}
H^{\psi \psi} & H^{\psi \lambda} \\
H^{\lambda \psi} & H^{\lambda \lambda} 
\end{array}
\right],$$
with $H_{\psi \psi}=E_g\{-\partial^2 cl(\theta;\mathbf{Y})/ \partial \psi \partial \psi  \}$ and $H_{\psi \lambda}=E_g\{-\partial^2 cl(\theta;\mathbf{Y})/ \partial \psi \partial \lambda  \}$.

\subsection{Composite likelihood inference in the likelihood paradigm}\label{section:CLR_evidence_function}

We propose that the composite likelihood and its corresponding set of all possible likelihood ratios can be used as a surrogate for the real likelihood ratios to provide pure likelihood inference for a given date set. For this, we need to prove that the composite likelihood functions have the two crucial performance properties possessed by real likelihood functions and some pseudo likelihoods (Section \ref{section:lik_paradigm}). Since composite likelihoods can be seen as misspecified likelihoods, we need to derive the robust adjustment factor defined in \citet{royall:2003}, so that the inference becomes robust against model misspecification. As a first condition, we need to determine whether \textit{the object of inference} is equal to \textit{the object of interest}, which can only be checked after the working model $f$ is chosen. 

In Theorem \ref{thm:composite_proof}, we show that composite likelihood functions, with the robust adjustment factor, have the two important performance properties of the likelihood paradigm \citep{royall:2003}.

\begin{thm}\label{thm:composite_proof}
\normalfont Assume $\mathbf{Y}=(Y_1,Y_2,...,Y_m)$ is a random vector from an unknown distribution $g(\mathbf{y})$. The parametric model $f(\theta; \mathbf{y})$ is chosen as the working model, with $\theta \in \Theta \subset \mathcal{R} $. 
Let $\theta_g$ be the (unique) minimizer of the composite Kullback-Leibler divergence between $f$ and $g$. 
Assume $\boldsymbol{Y_1},...,\boldsymbol{Y_n}$ is $n$ independent and identically distributed observations from the model $g(.)$. Under regularity conditions on the component log densities in Appendix A of the supplementary material, the following properties hold; (a) For any value $\theta \neq \theta_g$, the evidence will eventually support $\theta_g$ over $\theta$ by an arbitrarily large factor; $P_g (CL(\theta_g)/CL(\theta) \to \infty   \text{ as } n \to\infty )=1$. (b) In large samples, the probability of misleading evidence, as a function of $\theta$, is approximated by the bump function, $P_g \left( ( CL(\theta)/CL(\theta_g) )^{a/b}\geq k  \right) \to \Phi\left(-c/2-\log(k)/c\right),$ where $k>1$, $\Phi$ is the standard normal distribution function, $c$ is proportional to the distance between $\theta$ and $\theta_g$, $a=E_g\{\Delta_\theta u(\theta_g;\mathbf{Y})\}$ and $b=Var_g \{u(\theta_g;\mathbf{Y})\}$. The results can be extended to the case where $\theta$ is a fixed dimensional vector parameter. The proof is in Appendix A of the supplementary material. Note that we can substitute the $a$ and $b$ terms with the consistent estimates $\hat{a}=n^{-1} \Sigma_{i=1}^n u(\hat{\theta}_{CL};\mathbf{Y_i})$ and $\hat{b}=n^{-1} \Sigma_{i=1}^n (u(\hat{\theta}_{CL};\mathbf{Y_i}))^2$, where $\hat{\theta}_{CL}$ is the MCLE.
\end{thm}
Suppose $\theta \in \Theta \subset \mathcal{R}^d$ is partitioned as $\theta=(\psi,\lambda)$ and $\psi$ is a parameter of interest. It was shown in \citet{royall:2000} that the large-sample bound for the probability of misleading evidence, $\Phi(-(2\log k)^{1/2})$ holds for profile likelihoods. In Theorem \ref{thm:composite_profile_proof},  we show that the profile composite likelihood also has these two properties. 

\begin{thm}\label{thm:composite_profile_proof}
\normalfont 
Assume $\boldsymbol{Y}=(Y_1,Y_2,...,Y_m)$ is a random variable from an unknown distribution $g(y)$, the model $f(\theta;y)$ is the assumed model, with $\theta \in \theta \subset \mathcal{R}^2 $ partitioned as $\theta=(\psi,\lambda)$ and $\psi$ is a parameter of interest. Let $\boldsymbol{Y_1},...,\boldsymbol{Y_n}$ be $n$ independent and identically distributed observations from the model $g(.)$. Under regularity conditions on the component log densities in Appendix A of the supplementary material, the following properties hold.
\begin{enumerate}[label=(\alph*)]
\item  For any false value $\psi \neq \psi_g$, the evidence will eventually support $\psi_g$ over $\psi$ by an arbitrarily large factor;
\begin{eqnarray}\label{eq:property1_profile}
\frac{CL_p(\psi_g)}{CL_p(\psi)}&\to^p& \infty \quad \quad n\to\infty 
\end{eqnarray}
\item In large samples, the probability of misleading evidence, as a function of $\psi$, is approximated by the bump function,
\begin{eqnarray}\label{eq:property2_profile}
P_g\left\{ \left(\frac{CL_p(\psi)}{CL_p(\psi_g)}\right)^{a/b}\geq k\right\} &\to& \Phi\left(-\frac{c^*}{2}-\frac{\log(k)}{c^*}\right)
\end{eqnarray}
where $k>1$, $\Phi$ is the standard normal distribution function, $c^*=ca/b^{1/2}$, $c$ is proportional to the distance
between $\psi$ and $\psi_g$, $a=H^{\psi \psi}(\psi_g,\lambda_g)^{-1}$ and \\ $b=H^{\psi \psi} (\psi_g,\lambda_g)^{-1} G^{\psi \psi} (\psi_g,\lambda_g) H^{\psi \psi} (\psi_g,\lambda_g)^{-1} $.
\end{enumerate}
\begin{proof} \normalfont See Appendix A of the supplementary material.
\end{proof}
\end{thm}
The results can be extended to the case where $\psi$ and $\lambda$ are fixed dimensional vector parameters. Again, we substitute the $a$ and $b$  terms with consistent estimates. Note that the adjustment factor $a/b$ simplifies to $H^{\psi \psi}/G^{\psi \psi}$ since we assume $\psi$ is a scalar. This ratio is equal to the adjustment factor proposed by \citet{pace:2011}  in order to get a composite likelihood ratio test converging to a $\chi^2$ distribution instead of converging to $\sum \nu_i \chi^2_{(1)i}$, where the $\nu_i$'s are the eigenvalues of the matrix $(H^{\psi \psi})^{-1} G^{\psi \psi}$.

\subsection{Modelling correlated binary data using composite likelihoods}\label{section:modelling_corr_binary_data}

Consider a genetic association study where there are $N$ independent families with $n_i$ observations in the $i^{th}$ family, $i=1,...,N$. Let $\mathbf{Y_i}=(Y_{i1},...,Y_{in_i})$ be a binary response for the $i^{th}$ family, where  $Y_{ij}$ indicates whether the individual $j$ in the $i^{th}$ family has the trait or not ($Y_{ij}$ =1 or 0, respectively). Similarly the genotype data vector at a particular SNP is defined as $\mathbf{X_i}=(X_{i1},...,X_{in_i})$, where the SNP genotypes, $X \in \{ 0,1,2\}$, represent the number of minor alleles for a given SNP.  We look at the relative evidence for different values of the $OR$s for SNPs genome-wide or in a candidate region.  

In general, constructing a fully specified probabilistic model for correlated binary data is challenging. A joint probability mass function (pmf) for correlated binary variables was first proposed by \citet{bahadur:1961}, which involves writing the joint probabilities as functions of marginal probabilities and second and higher order correlations.  Although the Bahadur representation provides a tractable expression of a pmf, it has some limitations (\citet{bahadur:1961}, \citet[chap. 7]{molenberghs:2005}).  Other approaches for modelling the joint pmf for correlated binary data include constructing multivariate probit models or Dale models \citep{molenberghs:2005}. However, these are computationally intensive and hence intractable in high dimensional data.

Since evaluating the full likelihood is complicated, we construct a composite likelihood to model pedigree data and use the ratio of the composite likelihoods as our evidence function. We showed in Section \ref{section:CLR_evidence_function} that composite likelihood can be used as a surrogate for the real likelihood function for pure likelihood analysis and evidential interpretation assuming the object of interest and object of inference are the same.

The simplest composite likelihood to construct is from independent margins and is useful if one is interested only in marginal parameters \citep{varin:2011}.
Here, we are interested in the marginal parameter $\beta_1$ which is a $1^{st}$ order parameter. Thus, we choose a composite likelihood constructed from lower dimensional margins using the working independence assumption since they are easier to construct and they can be more robust to model misspecifications (\citet{jin:2010}, \citet{xu:2012}).
\begin{sloppypar}
Consider an underlying logistic regression model with an additive effect of the genotype on a binary response, $\log( p_{ij}/(1-p_{ij})) = \beta_0+\beta_1x_{ij}$, where $p_{ij}=P(Y_{ij}=1 | x_{ij})=E(Y_{ij} | x_{ij})$ is the marginal probability that the individual $j$ in the $i^{th}$ family has the disease trait given $x_{ij}=0,1$ or $2$.
The composite likelihood constructed under the working independence assumption is $CL_{ind}(\beta_0,\beta_1)=\Pi_{i=1}^N  \Pi_{j=1}^{n_i} P(Y_{ij}=y_{ij} \mid x_{ij})=\Pi_{i=1}^N \left( \Pi_{j=1}^{n_i} (p_{ij})^{y_{ij}}(1-p_{ij})^{1-y_{ij}}\right)$, where $p_{ij}=\exp(\beta_0+\beta_1x_{ij})/(1+\exp(\beta_0+\beta_1x_{ij})$. We can determine the profile composite likelihood $CL_{p}(\beta_1)=\max_{\beta_0}\{L(\beta_0,\beta_1)\}$ and compute the maximum profile composite likelihood estimate, $\hat{\beta_1}_{CL_{ind}}= \max_{\beta_1} \log CL_{ind}(\hat{\beta}_0(\beta_1),\beta_1)$. Note that we need to use the adjustment factor, $a/b$ in Eq. \eqref{eq:property2_profile} to the composite likelihood ratio. In the next section, we use a simulation study to investigate the implications of the theoretical results in Section \ref{section:CLR_evidence_function}.
\end{sloppypar}
\section{Simulation study}\label{section:simulation_section}
\subsection{Simulation design}\label{subsection:simulation_design_section}
\begin{figure} [H]
\centerline{ \rotatebox{0}{\resizebox{!}{10cm}{
  \includegraphics{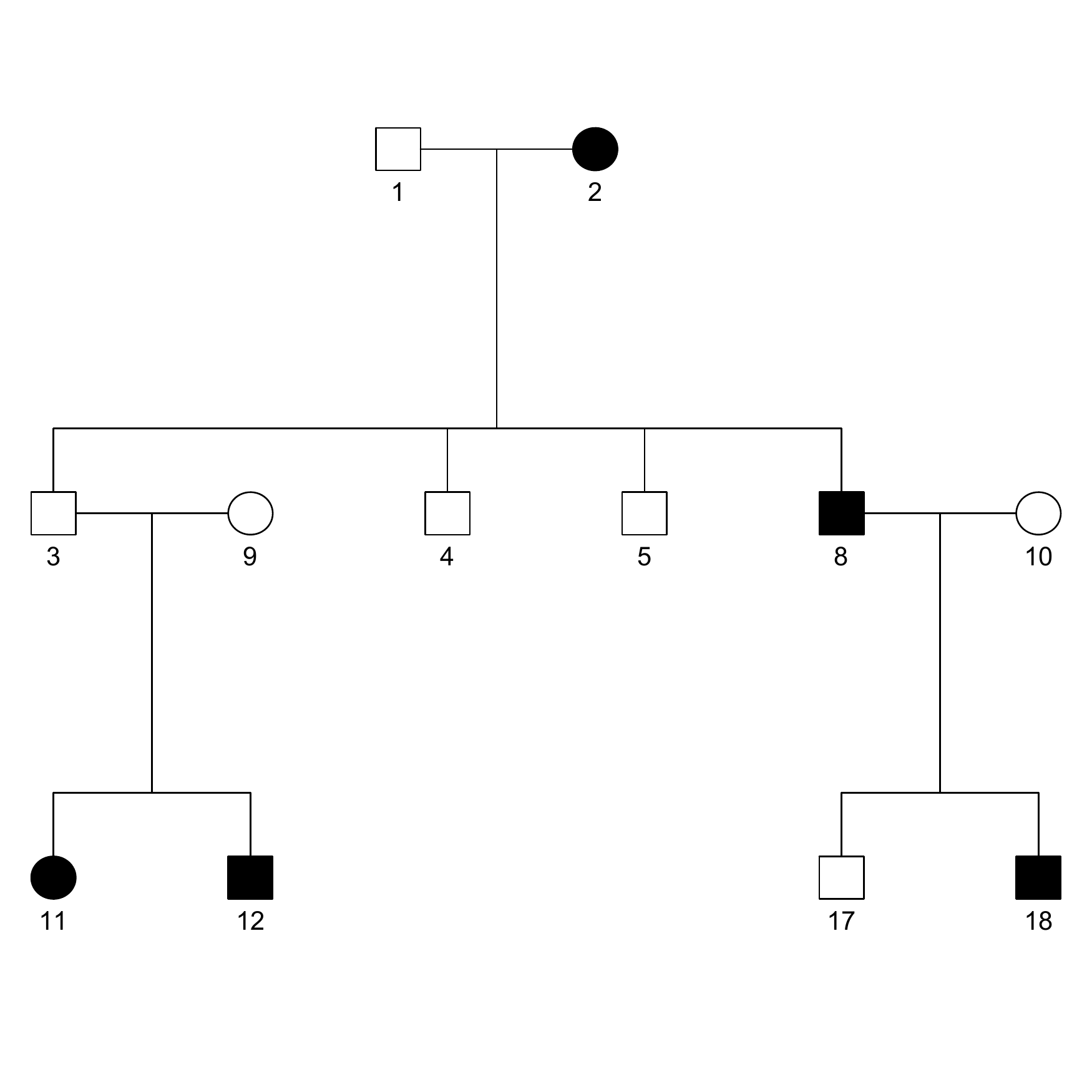}}}
   }
\caption{ Example of a generated family where there are 5 affected individuals. \label{fig:example_fam}}
\end{figure}

Consider a  family with 12 members, a proportion of whom are affected (eg. Figure \ref{fig:example_fam} ). We generate $N=30, 50, 100, 150, 200, 300$ and $500$ of such families with this structure to see how sample size affects the performance of our method.

We keep the regression parameters constant at $\beta_0=-2.38$ and $\beta_1=1.76$ throughout our simulations. With these values, the odds of disease when an individual does not carry the minor allele is 0.09 and the odds ratio is set large at 5.8.  
There are 5 values assumed for the dependence parameter: $\psi_1$ quantifies the dependence between siblings, $\psi_2$ quantifies the dependence between a parent and an offspring,  $\psi_3$ corresponds to the dependence between an aunt/uncle and a niece/nephew, $\psi_4$ quantifies the dependence between the grandparent and the grandchild and $\psi_5$ quantifies the dependence between cousins.  The values of the dependence parameters are chosen as  $\psi_1=3$, $\psi_2=2.5$, $\psi_3=2$, $\psi_4=1.5$, $\psi_5=1.2$.  We only assume positive dependence within pairs. See Table \ref{table:corr_oddsratio} and Appendix B of the supplementary material for the relationship between correlations within a binary pair and the odds ratio.

\begin{table}[htpb]
\caption{The relationship between correlations within a binary pair and odds ratio. \label{table:corr_oddsratio}}
\centering
 \begin{tabular}{l cc  cc cc cc}
 odds ratio &  &  &    \multicolumn{6}{c}{correlations}    \\
 $\psi$       &   &  &   $\delta_{ij | (0,0)}$ &  $\delta_{ij | (0,1)}$ &  $\delta_{ij | (1,1)}$ & $\delta_{ij | (0,2)}$ & $\delta_{ij | (1,2)}$ &$\delta_{ij | (2,2)}$ \\ 
\hline 
1.2    &   &  &   0.015 &  0.025 & 0.042 &  0.021 & 0.037 & 0.034 \\ 
3	&   &  &   0.120 & 0.155 & 0.253 & 0.099 & 0.197 & 0.222 \\ 
\end{tabular}
\bigskip

{\small where $\delta_{ij | (k,l)}= corr(Y_i,Y_j \mid x_i=k,x_j=l)$ and ($Y_i$,$Y_j$) is a binary pair.}
\end{table}
 
Genotype data ($\mathbf{X}$) for families  with a minor allele frequency of $0.20$ are generated using SIMLA \citep{schmidt:2005}. To generate the 12 dimensional correlated binary vector $\mathbf{Y}$ given $\mathbf{X}$, we use the method of \citet{emrich:1991}.  This method uses a discretised normal approach to  generate correlated binary variates with specified marginal logistic probabilities and pairwise correlations given genotype $\mathbf{X}$.  A detailed explanation of the data generation is given in Appendix B of the supplementary material. 

Our main purpose is to evaluate evidence about $\beta_1$ and determine if the procedure we propose will lead to valid inference. Our simulation must show: (1) that the maximum profile composite likelihood estimate of the parameter of interest converges to the true value as sample size increases to indicate that the object of inference is the same as the object of interest. That is, the composite likelihoods provide evidence about the true parameter (Eq.\eqref{eq:property1_profile} of Theorem \ref{thm:composite_profile_proof}), and (2) The probability of observing misleading evidence is described by the bump function (Eq.\eqref{eq:property2_profile} of Theorem \ref{thm:composite_profile_proof}).

For the composite likelihood constructed under the working independence assumption, we have $\theta=(\beta_0,\beta_1)$, where  $\beta_1$ is the parameter of interest and $\beta_0$ is the nuisance parameter. We follow the steps described in Appendix C of the supplementary material to find the maximum profile composite likelihood estimate of the parameter of interest, $\hat{\beta}_{1CL}$. We generate 10,000 simulated data sets and estimate the parameters by averaging over 10,000 maximum profile composite likelihood estimates, $\hat{\beta}_{1CL}=\sum_{j=1}^{10000} \hat{\beta}^{(j)}_{1CL_{p}}/10000$.
  
\begin{sloppypar}
To estimate the probability of misleading evidence (Eq.\eqref{eq:property2_profile} of Theorem \ref{thm:composite_profile_proof}), we first estimate the robust adjustment factor $a/b$. 
Recall  that $a=H^{\psi \psi}(\psi_g,\lambda_g)^{-1}$ and $b=H^{\psi \psi} (\psi_g,\lambda_g)^{-1} G^{\psi \psi} (\psi_g,\lambda_g) H^{\psi \psi} (\psi_g,\lambda_g)^{-1} $ where $\psi$ is the parameter of interest, $\lambda$ is the nuisance vector. Note that for $\theta=(\psi,\lambda)$, $G(\theta)=H(\theta) J(\theta)^{-1} H(\theta)$  with $H(\theta)=E_g\{-\partial^2 cl(\theta;\mathbf{Y})/ \partial \theta \partial \theta^T  \}$ and $J=E_g\{(\partial cl(\theta;\mathbf{Y})/ \partial \theta)(\partial cl(\theta;\mathbf{Y})/ \partial \theta)^T\}$. We estimate $J(\theta)$, using $\hat{J}(\hat{\theta})= 1/N \sum_{i=1}^N u(\hat{\theta}_{CL}, \mathbf{y_i})u(\hat{\theta}_{CL}, \mathbf{y_i})^T$,
where $u(\theta;\mathbf{y_i})$ are the elements of the composite score function, $\mathbf{y_i}$ is the observations vector, $\hat{\theta}$ is the global MCLEs of $\theta=(\beta_0,\beta_1)$, and $N$ is the sample size (number of families).

Then we estimate $H(\theta)$ by $\hat{H}(\hat{\theta})=\sum_{i=1}^N (\partial^2 cl(\theta;\mathbf{y_i})/ \partial \theta \partial \theta^T )/N$.
For the parameter of interest $\beta_1$,  $\hat{H}^{\beta_1 \beta_1}(\hat{\theta})$ and $\hat{G}^{\beta_1 \beta_1}(\hat{\theta})$ are the entries of the matrices $\hat{H}^{-1}(\hat{\theta})$  and $\hat{G}^{-1}$ that belong to the parameter $\beta_1$. Then, $\hat{a}/\hat{b}=  (\hat{H}^{\beta_1 \beta_1}(\hat{\theta})^{-1})/(\hat{H}^{\beta_1 \beta_1} (\hat{\theta})^{-1} G^{\beta_1 \beta_1} (\hat{\theta}) \hat{H}^{\beta_1 \beta_1} (\hat{\theta})^{-1}) =\hat{H}^{\beta_1 \beta_1}(\hat{\theta})/\hat{G}^{\beta_1 \beta_1}(\hat{\theta})$ since $\beta_1$ is a scalar.

To estimate the probability of misleading evidence for each simulated dataset, calculate the proportions of  the composite likelihood ratios with the robust adjustment factor that are greater than the pre-specified threshold, $(1/10000) \sum_{j=1}^{10000}  \mathcal{I} [  \{CL_p(\beta_1; \underset{\sim}{\mathbf{y}}^{(j)}) / CL_p(\beta_{1g};\underset{\sim}{\mathbf{y}}^{(j)} ) \}^{\hat{a}/\hat{b}} \geq k ]$, where $\underset{\sim}{\mathbf{y}}^{(j)}$ is the $j^{th}$ simulated dataset under the chosen model parameter $\beta_{1g}$, $\beta_1$ is a parameter value that is different than $\beta_{1g}$ and $\mathcal{I}$ is the indicator function.
\end{sloppypar}
\subsection{Simulation results}
The simulation results for determining whether the maximum profile composite likelihood estimates of $\beta_1$ converge to the true parameter value for sample sizes  $n=30, 50, 100, 150, 200, 300$ and $500$ are given in Table \ref{table:interest_inference_table}. We see that as $n$ increases, the composite likelihood approach provides consistent estimates for the true parameter value $\beta_1$. This ensures that \textit{the object of inference} is equal to \textit{the object of interest}; that is, the composite likelihood ratio is providing evidence about the true parameter of interest. 

\begin{table}[h!] \small
 \caption[The maximum profile composite likelihood estimates of $\beta_1$ for the simulation study.]{Simulation study. The maximum profile composite likelihood estimates of $\beta_1$, using the composite likelihood method and with different number of families ($n$) with $\beta_0=-2.38$, $\beta_1=1.76$. \label{table:interest_inference_table}}
 \vspace{0.5 cm}
 \centering
 \begin{tabular}{ r c ccccccccc  }
 $n$ & & 30 & 50 & 100 & 150 & 200 & 300 & 500     \\ 
\hline 
$\hat{\beta_1}_{CL_p}$  &&   1.798  & 1.782 & 1.772 & 1.767 & 1.766 &  1.762 & 1.762 \\ 
\end{tabular}
\end{table} 

\begin{figure}[H]
    \centering
    \begin{subfigure}[b]{0.45\textwidth}
        \includegraphics[width=\textwidth]{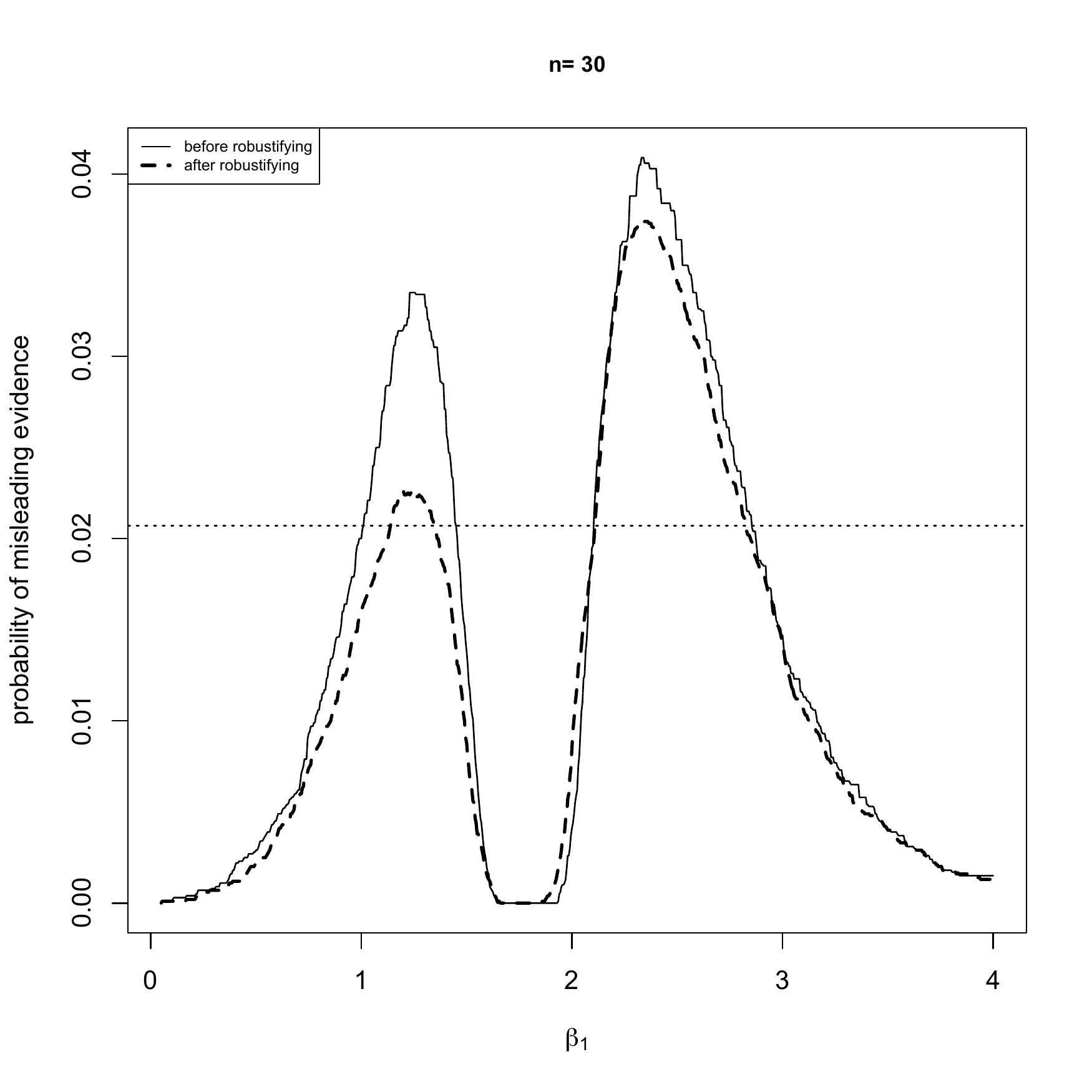}
        \caption{}
    \end{subfigure}
       \begin{subfigure}[b]{0.45\textwidth}
        \includegraphics[width=\textwidth]{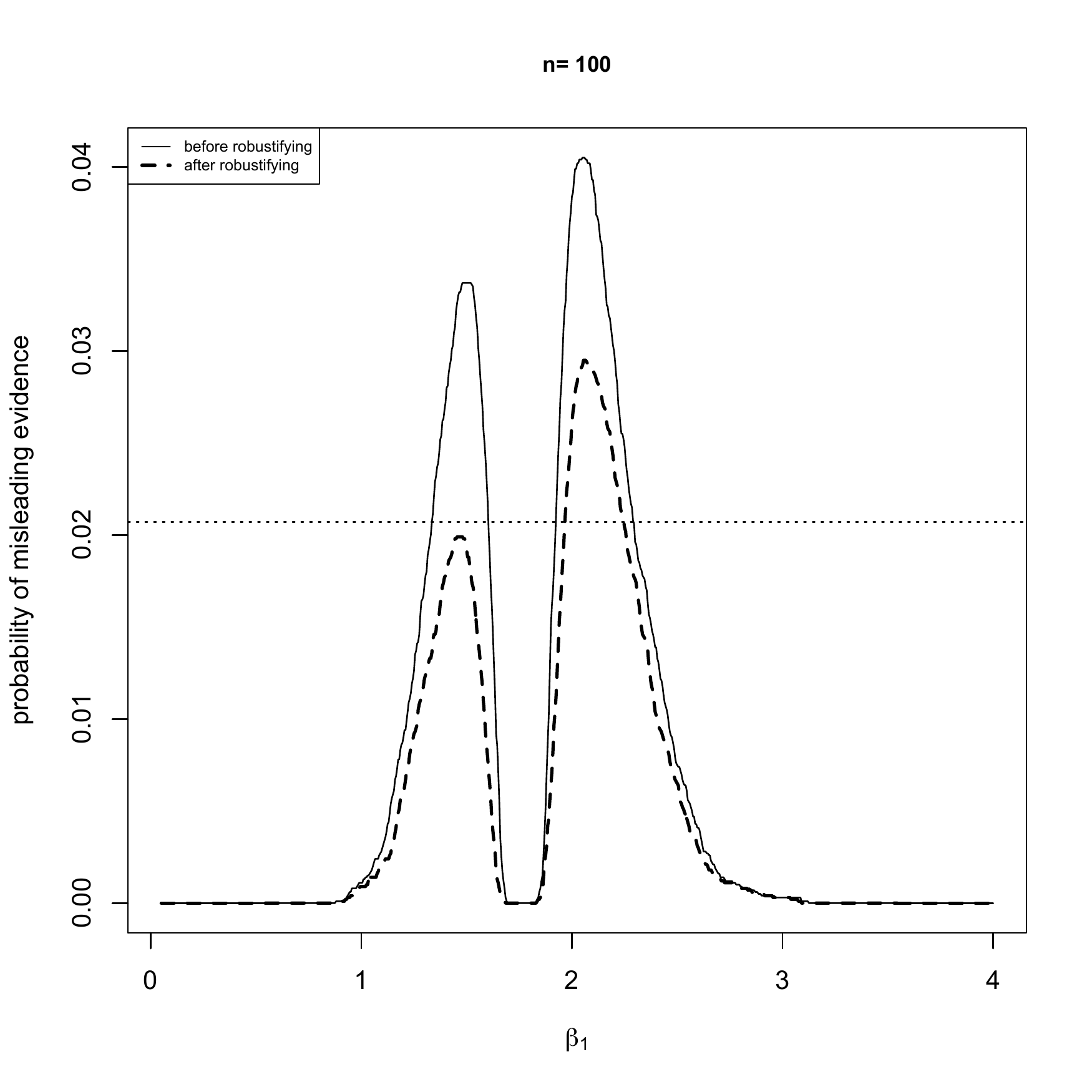}
        \caption{}
    \end{subfigure}
    \begin{subfigure}[b]{0.45\textwidth}
        \includegraphics[width=\textwidth]{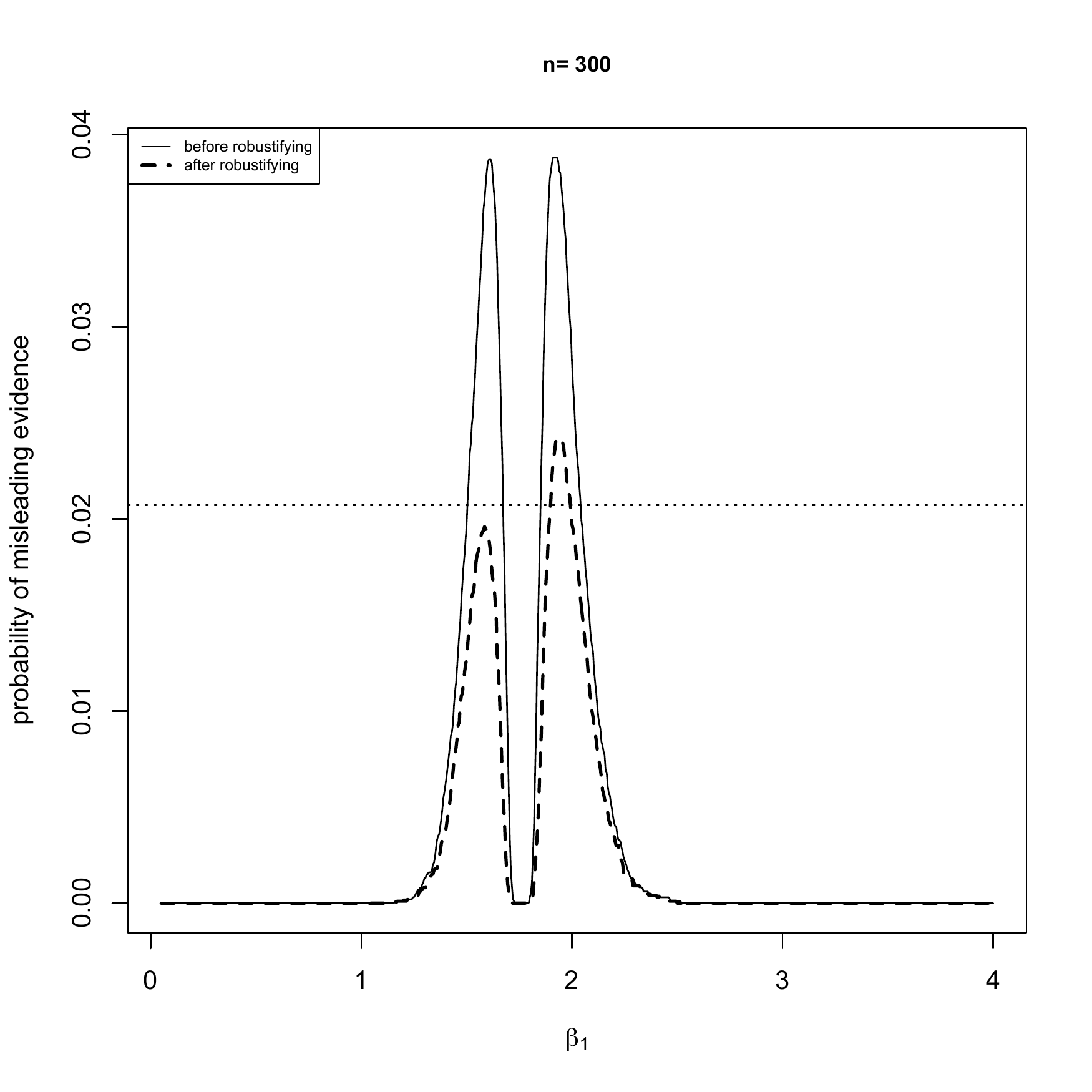}
        \caption{}
    \end{subfigure}
    \begin{subfigure}[b]{0.45\textwidth}
        \includegraphics[width=\textwidth]{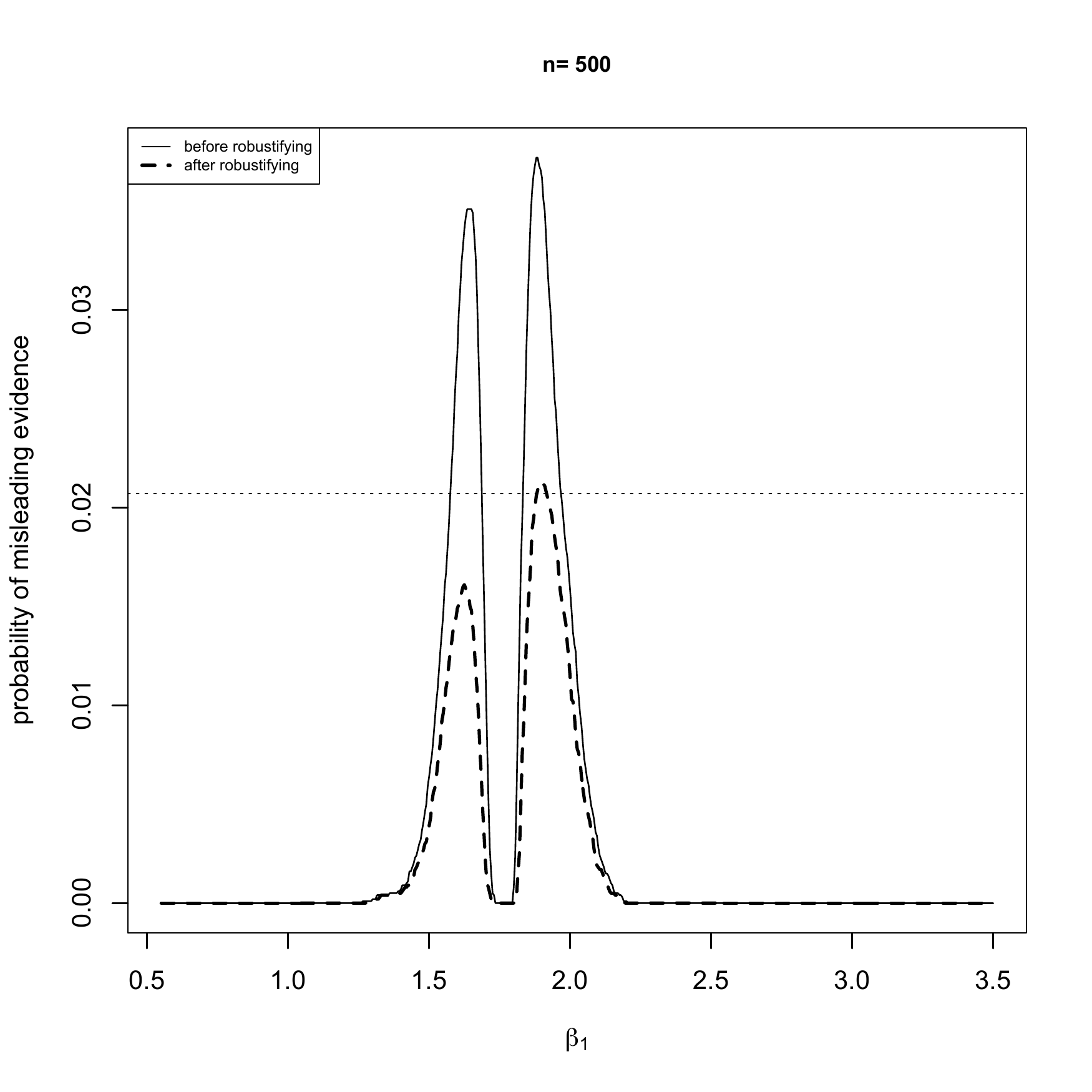}
        \caption{}
    \end{subfigure}
 \caption{Plots for the probability of misleading evidence before (- -) and after(--) robust adjustment for $n=30$, $n=100$,  $n=300$, $n=500$ with $\beta_0=-2.38$, $\beta_1=1.76$.}
\label{fig:simulation_bump}
\end{figure}

In Figure \ref{fig:simulation_bump}, we illustrate the behaviour of the probability of observing misleading evidence for $\beta_1$ for $n=$ 30, 100, 300 and 500. The solid curve indicates the probability of misleading evidence before the robust adjustment factor is applied and the dashed curve indicates this probability after the robust adjustment is applied. For illustration purposes, we chose $k=8$. As the theoretical results predict, after robust adjustment, the probability of observing misleading evidence is approximated by the bump function with increasing sample size. The bump function has the maximum value of $\Phi(-\sqrt{2 \log 8})=0.021$, indicated by the horizontal line in the figures.

In simulations where the number of individuals in families varies but only one type of relationship exists, e.g. only siblings, we looked at the performance of the likelihood ratios constructed from independent marginals and from pairwise marginals. We also considered the parameter that defines the relatedness, $\psi$, as the parameter of interest in the pairwise composite likelihood approach (Appendix D of the supplementary material). These simulations indicate that 1) both composite likelihood approaches provide valid evidence functions, 2) the composite likelihood approach from the independent marginals still performs well when there is more than one type of relatedness in the family, while the pairwise composite approach does not, 3) inference about the first order parameter $\beta_1$ is not affected by varying the value of the dependence parameter, and 4) if the parameter of interest is the second order parameter, then the pairwise likelihood approach is required to enable inference about the second order parameter (Appendix E of the supplementary material).

\section{Genetic Association Analysis of Reading Disorder in Families with Rolandic Epilepsy}

Rolandic Epilepsy (RE) is a neuro-developmental disorder characterized by centrotemporal sharp waves on EEG, focal seizures and a high frequency of comorbid speech and reading disorder (RD) in RE cases and their seizure unaffected family members. We conducted linkage analysis of RD in RE families and here we use our composite likelihood approach for analysis of genetic association at the Chromosome 1 RD linkage locus we identifed in the families (Chr 1: 209,727,257- 232,025,174) (\citet{strug:2012}). The data consists of 137 families and 1000 non-RD and non-RE control singletons. Some families are complex with up to 15 members. In total, there are 444 individuals in the RE families with 127 affected with RD.  All have been genotyped genomewide on the Illumina Human Core Exome array. At this locus there are 2087 genotyped SNPs analyzed for association.
  
We constructed the composite marginal likelihood under a working independence assumption with a robust adjustment factor, to correct for the misspecified model for correlated individuals and we assumed an underlying logistic regression model with an additive model for the SNP. The composite likelihood function with the robust adjustment factor, $a/b$, is $CL_{ind}(\beta_0,\beta_1)= \{ \Pi_{i=1}^N  \Pi_{j=1}^{n_j} \left( p_{ij} \right)^{y_{ij}} \left(1- p_{ij}  \right)^{1-y_{ij}} \}^{a/b}$ where $p_{ij}=\exp( \beta_0+\beta_1 x_{ij})/(1+\exp( \beta_0+\beta_1 x_{ij})$ and  $a/b=H^{\beta_1 \beta_1}(\beta_0,\beta_1)/G^{\beta_1 \beta_1}(\beta_0,\beta_1)$. Here the odds ratio (OR), $e^{\beta_1}$ is the interest parameter, and we plot the likelihood as a function of $e^{\beta_1}$ \citep{strug:2010}. Under the hypothesis of no association, the OR is equal to 1, and the OR is some value different from 1 under the alternative. Note that since $\beta_0$ is a nuisance parameter, we profile out the baseline odds, $e^{\beta_0}$, and use the profile composite likelihood $(CL_p)$ ratio with the robust adjustment factor as our evidence function, i.e. $( CL_p (e^{\beta_1}) /CL_p (1) )^{\hat{a}/\hat{b}}$.

\begin{figure}[H]
  \begin{minipage}[b]{0.50\linewidth}
    \centering
    \includegraphics[width=\linewidth]{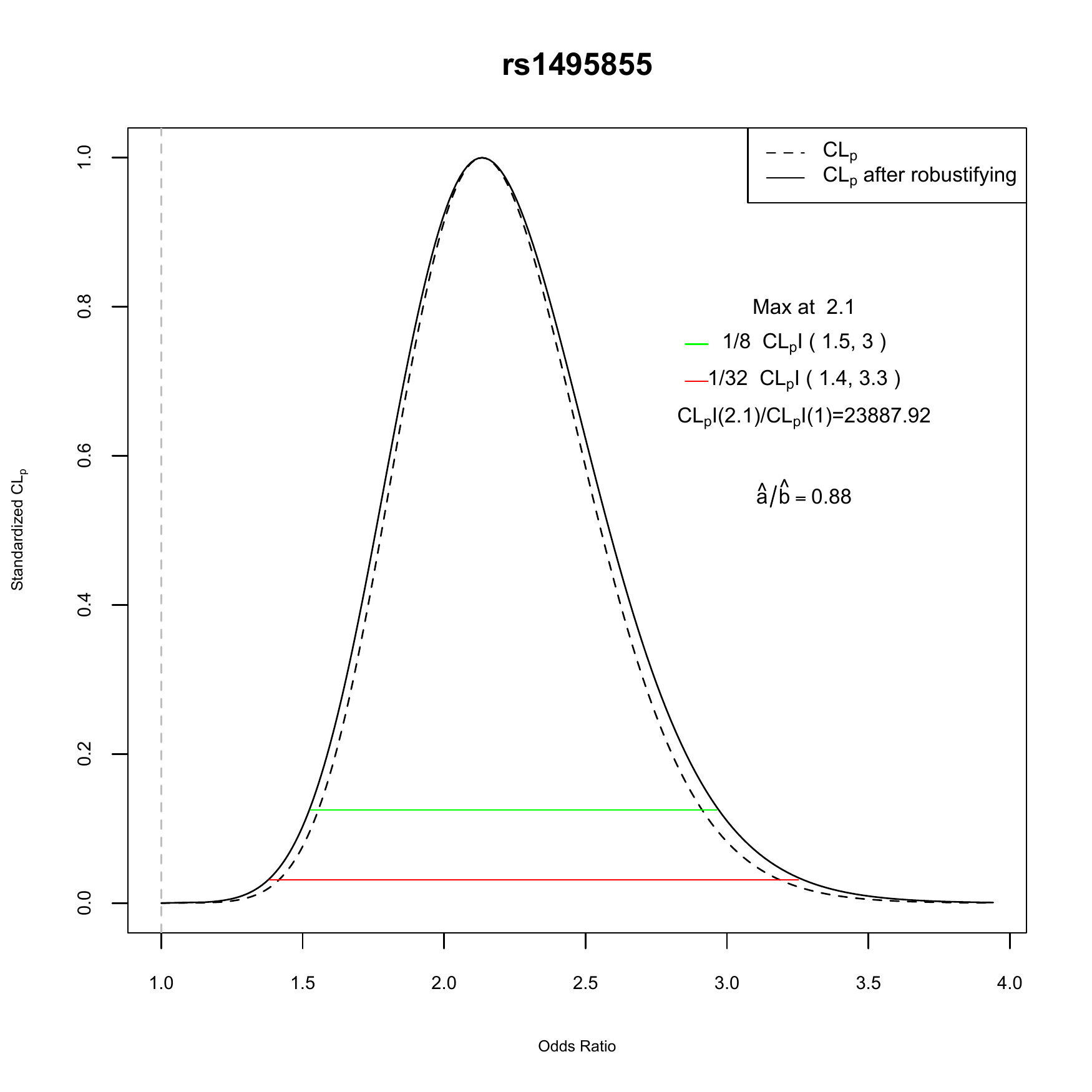}\\
    (a)
  \end{minipage}
  \begin{minipage}[b]{0.50\linewidth}
    \centering
    \includegraphics[width=\linewidth]{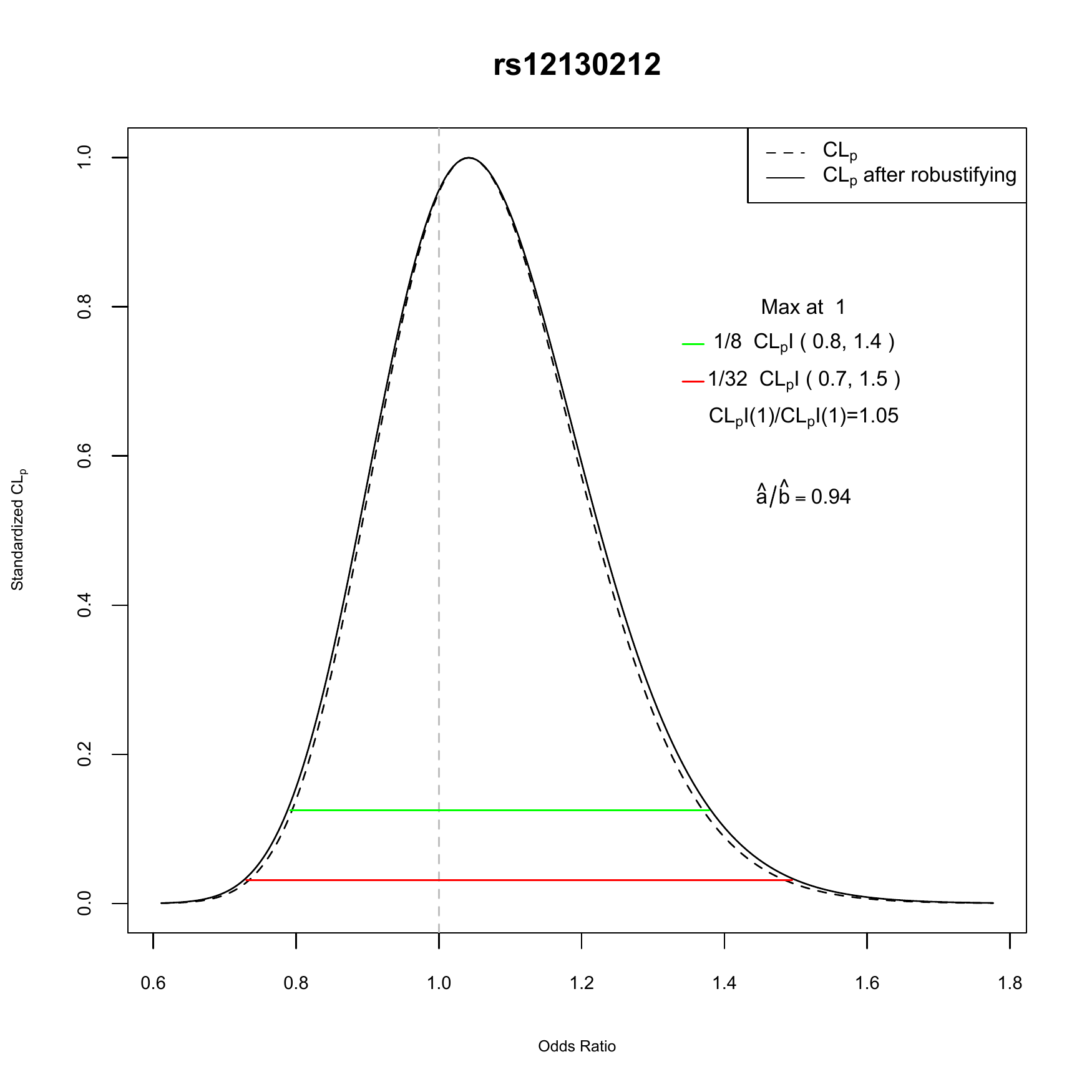}\\
    (b)
  \end{minipage}
\begin{minipage}[b]{0.50\linewidth}
    \centering
    \includegraphics[width=\linewidth]{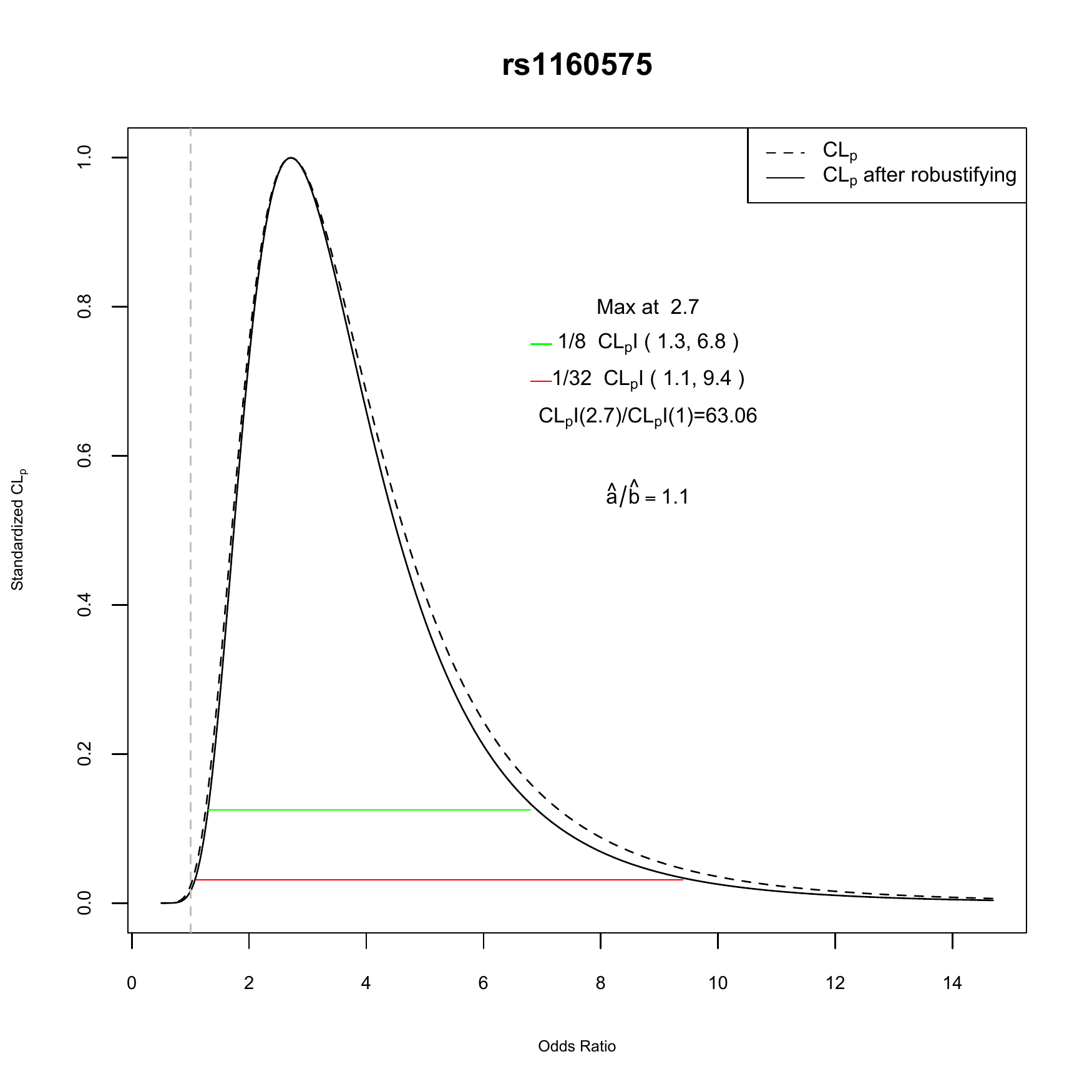}\\
    (c)
  \end{minipage}
\caption[Standardized $CL_p$ function for OR at two SNPs with 1/8 and 1/32 $CL_p$ intervals.]{Standardized $CL_p$ function for OR at three SNPs. 1/8 and 1/32 $CL_p$ intervals $(CL_pI)$ as well as the estimated robust adjustment factor $\hat{a}/\hat{b}$ are provided.}
\label{fig:3snps}
\end{figure}

In Figure \ref{fig:3snps}, we illustrate the $CL_p$ function for the OR for three SNPs: (a) a SNP, rs1495855, displaying association evidence, (b) a SNP, rs12130212, that does not show association evidence and (c) a SNP, rs1160575, with a low cell count in the $2 \times 3$ table (Table \ref{tab:snpskewed}). By plotting the $CL_p$ function, we can observe all the evidence about the association parameter $e^{\beta_1}$ that the data set provides.  

In Figure \ref{fig:3snps}(a), the ratio of any two points on the curve represent their relative support and the theoretical and simulation results ensure this interpretation is valid. The 1/8 $CL_p$ interval for the OR is 1.5 to 3. The OR values within this interval are consistent with the data at the level $k$=8, i.e. there are no other values outside this interval that are better supported than the values within the interval, by a factor of 8 \citep{royall:1997}.
We see that OR=1 is outside of the $1/8$ $CL_p$ interval. That tells us that there are some parameter values of the OR, for example the MLE, $ \hat{OR}_{mle}=2.1$ and nearby values, that are better supported than an OR=1 by a factor of greater than 8. The 1/32 $CL_p$ interval shows that an OR=1 is also not supported by the data at level $k$=32. The adjustment factor $\hat{a}/\hat{b}$ is 0.88, which is close to 1, suggesting that the composite likelihood is not too discrepant from the true likelihood. This is due to the fact that most individuals in our data are unrelated with the 1000 singletons included in the analysis.

In Figure \ref{fig:3snps}(b), we can see that both 1/8 and 1/32 $CL_p$ intervals include OR=1 as a plausible value. This indicates us that there is no value that the data supports over OR=1 by a factor of 8 or more. In Figure \ref{fig:3snps}(c), the $CL_p$ is skewed suggesting there is sparsity in the data (see Table \ref{tab:snpskewed}).

\begin{table}[H]
  \centering
  \caption{Distribution of data at SNP rs1160575 }
    \scalebox{0.9}{
    \begin{tabular}{rcrrr}
          &       & \multicolumn{3}{c}{disease status} \\   \cline{3-5}
          & \# of minor allele & no    & yes   & total \\   \cline{1-5}
    \multirow{3}[0]{*}{SNP} & 0     & 1160  & 121   & 1281 \\
          & 1     & 153   & 6     & 159 \\
          & 2     & 4     & 0     & 4 \\    \cline{1-5}
    total &       & 1317  & 127   & 1444 \\
    \end{tabular} }
  \label{tab:snpskewed}
\end{table}

\begin{figure} [H]
\centerline{ \rotatebox{0}{\resizebox{!}{15cm}{
  \includegraphics{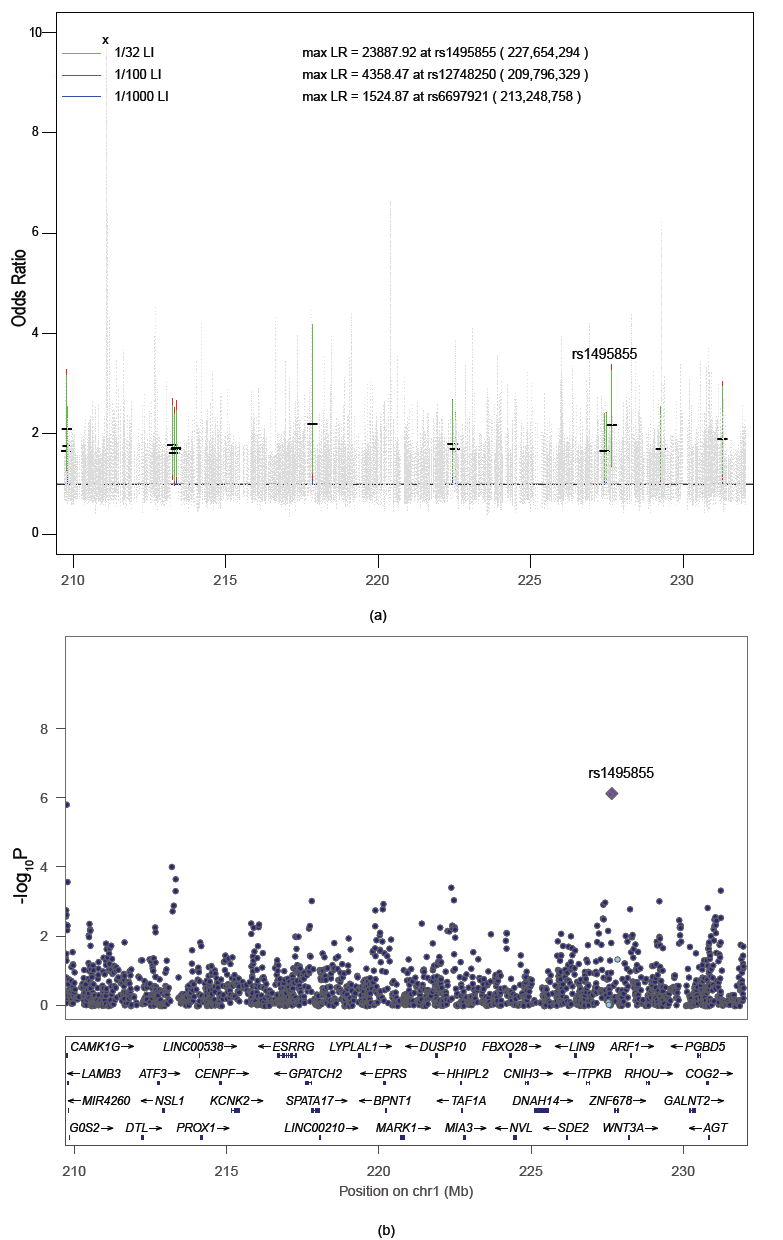}}}
   }
  \caption{Evidential analysis of association between SNPs at chromosome 1 and RD (0 or 1) using a composite likelihood from independent margins where the margins are logistic regression models with genotypes coded additively, with a robust adjustment (a).  Analysis of the data using a GEE approach with an independent correlation structure; y-axis provides $\log_{10}P$-values (b).} 
  \label{fig:many_snps}
\end{figure}

In genetic association studies,  one often needs to evaluate many SNPs in a region or genome-wide. Plotting hundreds of individual likelihood functions corresponding to each SNP may not be practical. Instead, a single plot that represents the association information in a region of interest can be displayed \citep{strug:2010} and followed up by individual likelihood plots at a small number of markers of interest. Figure \ref{fig:many_snps}(a) presents the $CL_p$ intervals for  2087 SNPs under the Chromosome 1 linkage peak from 1444 individuals. We display the SNPs by base pair position on the x-axis and OR on the y-axis. The vertical lines for each SNP on the x-axis represent the 1/$k$ $CL_p$ intervals where $k=32$, $100$ and $1000$.  Only the SNPs whose 1/100 $CL_p$ intervals do not include the OR=1 as a plausible value are displayed in colour and noted as providing strong evidence for ORs different from 1 for a given $k$. The $CL_p$ intervals displayed in grey include OR=1 (horizontal line at OR=1) as a plausible value so the SNPs that produced these intervals are not concluded to display strong evidence for association. If the $1/k$ $CL_p$ interval does not include OR=1, then it will be coloured in green, red or navy blue for $k=32, 100$ and $1000$ respectively. For SNP rs1495855, the $1/k$ $CL_p$ interval for $k=32, 100$ and $1000$ do not include OR=1 as a plausible value, indicating evidence of an association between this SNP and RD at the level $k>1000$. Note that the $1/32$ $CL_p$ interval for rs1495855 (the green portion) is the same as provided in Figure \ref{fig:3snps}(a).
The longest grey $CL_p$ interval at the left of the plot (marked with an \textbf{x} in the plot) corresponds to the SNP in Figure \ref{fig:3snps}(c). Therefore, Figure \ref{fig:many_snps}(a) also provides information about the shape of the likelihood function for a given SNP.

 The small horizontal tick on each $CL_p$ interval represents the MLE for the OR at the SNPs that were found to be associated with RD for some $k$. The max $CL_p$ ratios, calculated by $[CL_p(\hat{OR}_{mle})/CL_p(1)]^{\hat{a}/\hat{b}}$, for the three SNPs where the strength of evidence for association is the largest are also provided on the plot.  The SNP rs1495855 provided the largest likelihood ratio with an MLE for the OR=2.1.

Using GEE with an independent correlation structure to assess association provides results that are consistent with the composite likelihood approach. That is,  all of the SNPs that provide $CL_p$ ratios for ORs that are better supported than an OR=1 by a factor of greater than 100 are among the ones that produce $P$-values$<$0.01 in the GEE approach. Moreover, the three SNPs that have the maximum likelihood ratios also have the smallest $P$-values in the GEE analysis, i.e. $P$-values for SNPs rs1495855, rs12748250 and rs6697921 are $<0.000001$, $<0.00001$ and $<0.0001$ respectively.  Figure \ref{fig:many_snps} compares the GEE analysis results with the evidential analysis results. In Figure \ref{fig:many_snps}(b), the SNPs by base pair position are displayed on the x-axis as in Figure \ref{fig:many_snps}(a) and the $\log_{10}P$-values for the corresponding SNPs from the GEE analysis are displayed on the y-axis. The smallest $P$-value corresponds to SNP rs1495855 and the $OR$ estimated from the GEE analysis is $2.1$.  Figure \ref{fig:many_snps}(b) only indicates that the probability of observing a result this extreme or more is unlikely if the true OR is 1.

\subsection{Multiple Hypothesis Testing Adjustments}

Until now we have not considered multiple hypothesis testing. The probability of observing a $LR>k$ at a single SNP if $OR=1$ is true is bounded (Eq. \eqref{eq:property2_profile}). But if one aims to have the probability across all SNPs bounded, there are some alternative considerations. Let $H_0$ be the hypothesis that none of the $N$ SNPs are associated with the trait. To control the family-wise error rate (FWER), which is the probability of at least one $LR>k$ among $N$ hypotheses when $H_0: OR=1$ is true, let $M_0(n,N,k)$ be the FWER where $n$ is the sample size and $k$ is the criterion for the measure of evidence. Then, 
\begin{eqnarray*} 
FWER=M_0(n,N,k) &=&   P_0 \left( (LR_1 \geq k) \cup (LR_2 \geq k) \cup ...\cup (LR_N \geq k) \right) \\
&\leq & P_0 (LR_1 \geq k) +P_0 (LR_2 \geq k)+...+P_0 (LR_N \geq k) \\
&=& \sum_{j=1}^N M_0^{(j)} (n,k)  \quad \quad \quad j=1,...,N \text{   SNPs }      
\end{eqnarray*}

where $M_0^{(j)} (n,k)$ is the probability of observing misleading evidence at the $j$-th SNP for two simple hypotheses for the OR. $M_0^{(j)} (n,k)$ is a planning probability and so generally, $H_0:OR=1$, and $H_1$ is the OR that is the minimum clinically important difference, since for larger ORs, $M_0$ is smaller \citep{strug:2006b}. 

For planning purposes $M_0^{(j)} (n,k)$ are the same for all $j$. Thus, a conservative upper bound on the FWER is $N M_0 (n,k)$. Since $M_0(n,k)$ is usually very small for any given SNP,  it may provide a reasonable upper bound for the FWER. Otherwise, increasing the sample size can dramatically lower the FWER \citep{strug:2006a}, suggesting sample size is an adjustment for multiple hypothesis testing.

We simulated $10^5$ replicates under no association ($OR=1$). Let $\underset{\sim}{\mathbf{y}}^{(j)}$ be the $j^{th}$ simulated phenotype under $H_0$ given the corresponding genotype, $CL_p( OR;\underset{\sim}{\mathbf{y}}^{(j)})$ is the profile composite likelihoods evaluated at a chosen OR, and $\mathcal{I}$ is an indicator function. Then we evaluate the likelihood ratio for $OR=2$ and $OR=2.5$ versus $OR=1$ and we estimate $M_0(n,k)$ using $(1/10^5) \sum_{j=1}^{10^5}  \mathcal{I}  [ \{CL_p(OR; \underset{\sim}{\mathbf{y}}^{(j)}) / CL_p( 1;\underset{\sim}{\mathbf{y}}^{(j)} ) \}^{\hat{a}/\hat{b}}  \geq k ]$. The effective number of independent tests in the set of 2087 dependent markers was estimated as 1413 \citep{li:2012}. Thus, $N=1413$ is used in the calculation of the upper bound for the $FWER$. 

We calculate the upper bound on the FWER for different choices of criterion $k$ ($k=32, 64, 100$ and $1000$). For $k=1000$, the upper bound for the FWER is 0.3250 at $OR=2$. This tells us that the probability of observing at least one $LR>k$ among 1413 SNPs when none of the SNPs are associated with the trait is bounded by 0.3250. A lower upper bound would be preferable, however, although we know that this is a crude upper bound and may not provide a good estimate of the true FWER, it is straightforward to estimate.

When feasible,  the best approach to decrease the $M_0(n,k)$ and consequently the FWER is by increasing the sample size or equivalently, replicating the result in an independent sample. This approach also reduces the probability of weak evidence whereas increasing $k$ for a fixed $n$ increases weak evidence. To see how increasing the sample size will effect the upper bound on the $FWER$, we simulated a data set based on the original data structure, but where the number of families is doubled from 444 to 888. We see in Table \ref{table:FWER_sim_double}  that the upper bound on the FWER is 0.1837 and is considerably lower with greater sample size, where we now have 1888 individuals, instead of 1444.

\begin{table} [H]
  \caption{Upper bound for FWER for $10^5 $replicates when the sample size is 1888.}
    	\label{table:FWER_sim_double}
\begin{center}
    \begin{tabular}{lccccc} 
    \hline
     \vspace{0.1cm}
     FWER $\leq$ N $M_0(1888,k))$ &  $k=32$  &$ k=64$  & $k=100$ & $k=1000$   \vspace{0.1cm} \\ \hline
     \vspace{0.1cm}
  \quad \quad \quad      OR=2 &  1  & 1   & 1  & 0.1837    \vspace{0.1cm} \\    \hline
  \vspace{0.1cm}
 \quad \quad \quad      OR=2.5 &  1  & 0.6076  & 0.4098    & 0.0706  \\    \hline
 \vspace{0.1cm}
\end{tabular}
\end{center}
\end{table}

\section{Summary}

We have developed an alternative approach to the analysis of genetic association for correlated (family) data using the pure likelihood paradigm. The likelihood paradigm provides a full likelihood solution that enables more comprehensive inference than null hypothesis significance testing.  Due to complex dependencies in family data, constructing a fully specified probabilistic model for a binary trait is challenging. Therefore, we considered working with composite likelihoods for modelling this type of data, which has also been considered in a frequentist context (\citet{cessie:1994}, \citet{kuk:2000}, \citet{zhao:2005}, \citet{zi:2009}, \citet{he:2011}). We showed that LRs from composite likelihoods, with a robust adjustment,  are valid statistical evidence functions in the likelihood paradigm. They have the two required performance properties of an evidence function, assuming the object of inference is equal to the object of interest, that enable the measurement of evidence strength by comparing likelihoods for any two values of an interest parameter. The robust adjustment on the composite likelihood is necessary even though the likelihood objects in the composite likelihood are correctly specified, since multiplying them to construct the composite likelihood does not in general lead to a probability density function.

If one is interested in marginal parameters (e.g. $OR=e^{\beta_1}$), we proposed constructing composite likelihoods from independent marginals when we have complex family structures. (Working with these independent likelihoods reduces the computations considerably.) Using simulation, we also examined the use of composite likelihoods for a logistic regression model with an additive effect on the marginal binary response, and we show that this choice of composite likelihood offers reliable inference as well. The composite likelihood approach contributes additional information by providing a full likelihood solution that can complement frequentist GEE analysis, and is more feasible to implement over generalized mixed models. 

We applied the composite likelihood method to the analysis of genetic association on Chromosome 1 at the RD linkage locus in RE families. We found that rs1495855 provided large likelihood ratios for ORs near 2 versus OR=1. We observed an MLE of OR=2.1 and almost 24000 times greater evidence for OR near 2 versus OR=1. The $1/1000$ likelihood interval is $(1.16,3.89)$ (not shown on Figure \ref{fig:many_snps}(a)). Even the values around 1.2 are better supported over OR=1 by a factor of 1000. GEE analysis also supported evidence for association at this variant. Lastly, we discussed how FWER control is achieved in the context of this paradigm, and showed that indeed the probability of observing a misleading result across the 2089 SNPs at even $k>1000$ was actually quite high and a replication sample would be needed to decrease the FWER.

A limitation of this approach is that it may not be optimal in small samples since the performance properties for incorrect models (e.g. composite likelihoods) rely on large sample results.  Future work will determine an efficient solution for small sample adjustments, potentially using a Jacknife variance estimate as was done in small sample correction methods for GEE (\citet{paik:1988}, \citet{lipsitz:1990}). Another challenge is when interest lies in higher order parameters, like the correlation parameter. In this case, composite likelihoods that are composed of more complex marginals are required for pure likelihood inference. This makes the computations more difficult and leads to longer computational time. Finding a working model where the object of interest is equal to the object of inference may also be challenging, which is critical when working with incorrect models in any paradigm.

In conclusion, we have provided a composite likelihood approach for the analysis of genetic association in family data using the likelihood paradigm. Our method is practical, efficient and easy to implement and provides a reliable evidence function when the real likelihoods are intractable or impractical to construct. 

\section*{Software}
\label{sec5}

Software in the form of R code, together with the simulated data set is available from the authors upon request.

\section*{Acknowledgements}

We thank Deb K. Pal who coordinated the Rolandic Epilepsy study and its data collection.  We thank all the families from around the world who contributed to the study.  We acknowledge the contributions of our funders, the Canadian Institutes of Health Research (201503MOP-342469 to LJS and DKP) and the Natural Sciences and Engineering Research Council of Canada (2015-03742  to LJS). \\
{\it Conflict of Interest}: None declared.

\bibliographystyle{asa}
\bibliography{our_refs}
\newpage
\section*{Supplemental Material} 

 \bigskip

\appendix

\section{Proofs of Theorem 1 and Theorem 2 }\label{App:Appendix_proof}

\subsection{Regularity Conditions}
The regularity conditions (A1-A6) are provided in \citep*[p.245]{knight:2000}. Let $l(\theta;y)=\log f(y;\theta)$ and let $l_\theta(\theta;y), l_{\theta\theta}(\theta;y)$ and $l_{\theta\theta\theta}(\theta;y)$ be the first three partial derivatives of $l(\theta;y)$ with respect to $\theta$.  These conditions apply on the component log densities of a composite likelihood.

\begin{enumerate}
\item[$A1.$] The parameter space $\Theta$ is an open subset of the real line.
\item[$A2.$] The set $A=\{y:f(x;\theta)>0\}$ does not depend on $\theta$.
\item[$A3.$] $f(y;\theta)$ is three times continuously differentiable with respect to $\theta$ for all $y$ in $A$.
\item[$A4.$] $E[ l_\theta(\theta;y)]=0$ for all $\theta$ and $Var[l_\theta(\theta;y)]=I(\theta)$ where $0<I(\theta)<\infty$ for all $\theta$.
\item[$A5.$] $E [l_{\theta \theta}(\theta;y)]=-J(\theta)$ where $0<J(\theta)<\infty$ for all $\theta$.
\item[$A6.$] For each $\theta$ and $\delta>0$, $|l_{\theta \theta \theta}(t;y) \leq M(x) |$ for $|\theta-t| \leq \delta$ where $E_\theta[M(Y_i)] < \infty$
\item[$A7.$] There exists a unique point $\theta_g \in \Theta$ which minimizes the composite Kullback-Lebler divergence in Eq. \ref{eq:KL_divergence} \citep{xu:2012}.
\begin{eqnarray} \label{eq:KL_divergence}
K(g : f ; \theta)&=&\sum_{k=1}^K E_g \left[ \log g(Y \in A_k) -\log f(Y \in A_k;\theta) \right] w_k.
\end{eqnarray}
\end{enumerate}
Condition $A4$ changes when there is model misspecification (e.g. setting up wrong marginal or conditional densities in composite likelihoods), e.g. $E_g[ l_\theta(\theta;y)]=0$ only for $\theta=\theta_g$, where the expectation is taken under the correct (unknown) model $g$. 

It can be deduced that the mean and variance of the log likelihood derivatives,  $l_\theta(\theta;y), l_{\theta\theta}(\theta;y)$ and $l_{\theta\theta\theta}$ are of order $O(n)$. The higher order derivatives are, in general, of order $O_p(n)$ \citep[p.88]{severini:2000}.

According to \citet[p.106]{severini:2000}, sufficient conditions for the consistency of the MLE for regular models are:
\begin{enumerate}
\item $\Theta$ is a compact subset of $\mathcal{R}^d$.
\item $\sup_{\theta \in \Theta} \left| n^{-1} l(\theta)-n^{-1} E\{\l(\theta) \} \right| \to^p 0$ as $n \to \infty$.
\end{enumerate}
We need the second condition to hold on the component log densities. Let $l_i$ be the $i^{th}$ component log density in the composite likelihood with $i=1,..,d$, such that $cl(\theta)=\sum_{i=1}^d n^{-1} l_i(\theta) $ then
\begin{eqnarray} \label{eq:consistency_reg} \nonumber
 \left| \sum_{i=1}^d n^{-1} l_i(\theta)-\sum_{i=1}^d n^{-1} E\{\l_i(\theta) \} \right|&\leq &\sum_{i=1}^d  \left| n^{-1} l_i(\theta)-n^{-1} E\{\l_i(\theta)\} \right|  \\ \nonumber
 \sup_{\theta \in \Theta} \left| \sum_{i=1}^d n^{-1} l_i(\theta)-\sum_{i=1}^d n^{-1} E\{\l_i(\theta) \} \right|&\leq &  \sup_{\theta \in \Theta} \sum_{i=1}^d  \left| n^{-1} l_i(\theta)-n^{-1} E\{\l_i(\theta)\} \right|  \\   
 &= &  \sum_{i=1}^d  \sup_{\theta \in \Theta}   \left| n^{-1} l_i(\theta)-n^{-1} E\{\l_i(\theta) \} \right|  
\end{eqnarray}
If each component in Eq.\eqref{eq:consistency_reg} goes to 0 in probability, then the term on the left side goes to 0 in probability.
Furthermore, see \citep*{xu:2012} for a more detailed regularity conditions that are needed for the consistency of the maximum composite likelihood estimator.

\subsection{Proof of Theorem 1}
\begin{enumerate}[label=(\alph*)]
\item We want to show $P_g \{CL(\theta_g)/CL(\theta) \to  \infty   \text{ as } n \to\infty \}=1$.
The composite likelihood function for $n$ observations is  $CL(\theta)=\Pi_{i=1}^n CL(\theta;\mathbf{y}_i)$. Let $R_n=\Pi_{i=1}^n CL(\theta_g; \mathbf{y}_i) / \Pi_{i=1}^n CL(\theta; \mathbf{y}_i)$. We want $R_n \rightarrow \infty$. 

Let $cl(\theta;\mathbf{y})=\log CL(\theta;\mathbf{y})$.% We take $(1/n)^{th}$ power then $log$ of this quantity, i.e. $\log(R_n)^{1/n}$;
\begin{eqnarray} \label{eq:slln}  \nonumber
 \log \left(\frac{\Pi_{i=1}^n CL(\theta_g; \mathbf{y}_i)} {\Pi_{i=1}^n CL(\theta; \mathbf{y}_i)} \right)^{1/n}&=&\frac{1}{n} \left( \sum_{i=1}^n cl (\theta_g;\mathbf{y}_i) - \sum_{i=1}^n cl(\theta;y_i)  \right) \\ \nonumber
 &=&\frac{1}{n} \left( \sum_{k=1}^K w_k \left\{  \sum_{i=1}^n \log(f(\mathbf{y}_i \in A_k;\theta_g))  \right. \right.\\ \nonumber
 && \quad  \quad \quad \left. \left. - \sum_{i=1}^n \log(f(\mathbf{y}_i \in A_k;\theta))  \right\}        \right) \\   \nonumber
 & \rightarrow^{a.s} & \sum_{k=1}^K w_k \left\{ \mathbf{E}_{g} \left[ \log(f(\mathbf{Y} \in A_k;\theta_g)) \right. \right.\\
 && \quad \quad \quad  \left. \left.  -\log(f(\mathbf{Y} \in A_k; \theta))\right] \right\} \\ \nonumber
 &>& 0
\end{eqnarray}
since $w_k$'s are positive and $\theta_g$ minimizes the $K(g:f;\theta)$. \eqref{eq:slln} is true by the Strong Law of Large Numbers. Since $(1/n)\log R_n \rightarrow c > 0$, where $c$ is a finite positive number, then $R_n=  \Pi_{i=1}^n CL(\theta_g,\mathbf{y}_i) / \Pi_{i=1}^n CL(\theta,\mathbf{y}_i) \to \infty$

\item  We want $P_g\{ [CL(\theta)/CL(\theta_g) ]^{a/b}\geq k)\} \to \Phi(-c/2-\log(k)/c)$ where $c$ is proportional to the distance
between $\theta$ and $\theta_g$. Note that $CL(\theta;\mathbf{y})=\Pi_{k=1}^K f(\mathbf{y} \in \mathcal{A}_k;\theta)^{\omega_k}$ and $cl(\theta;\mathbf{y})=\log CL(\theta;y)=\sum_{k=1}^K \omega_k \log f( \mathbf{y} \in \mathcal{A}_k;\theta)$.\\
\textit{Composite score function:} $u(\theta;\mathbf{y})= \nabla_{\theta} cl(\theta;\mathbf{y})=\sum_{k=1}^K \omega_j \nabla \log f(\mathbf{y} \in \mathcal{A}_k ; \theta) $\\
The sensitivity matrix under the correct model:
$$H(\theta)=\mathbf{E}_g\left( -\nabla_{\theta} u(\theta; \mathbf{Y})  \right) =\int -\nabla_{\theta} u(\theta;\mathbf{y}) g(\mathbf{y}) d\mathbf{y}    $$
The variability matrix under the correct model:
$$J(\theta)=var_g (u(\theta;\mathbf{Y}))$$
The Godambe information matrix (Godambe,  1960) under the correct model is $$G(\theta)=H(\theta)J(\theta)^{-1}H(\theta). $$
Then for $n$ independent and identically distributed observations $\boldsymbol{Y_1},...,\boldsymbol{Y_n}$ from the model $g(.)$, as $n \rightarrow \infty$, under the regularity conditions, we have;
\begin{eqnarray} \label{eq:score}
\frac {\sum_{i=1}^n u(\theta_g;\mathbf{y_i})}{\sqrt{n}} & \rightarrow^d N\left(0,J(\theta_g)\right) \quad \quad \quad \text{since } E_g(u(\theta_g;\mathbf{Y}))=0
\end{eqnarray}
\begin{eqnarray} \label{eq:hessian}
\frac {\sum_{i=1}^n \nabla_{\theta} u(\theta_g;\mathbf{y_i})}{n} & \rightarrow -H(\theta_g)
\end{eqnarray}
where $\theta_g$ is the (unique) minimizer of the composite Kullback-Leibler divergence between $f$ and $g$. 
Let $\theta=\theta_g+c/\sqrt{n}$, then the Taylor expansion of the log composite likelihood around $\theta_g$ is;

\begin{align} \label{eq:bump_function1}  \nonumber
cl(\theta)-cl(\theta_g) &= u(\theta_g)(\theta-\theta_g)+
\nabla_{\theta} u(\theta_g) \frac{(\theta-\theta_g)^2}{2}+O_p(n^{-1/2}) \\ \nonumber 
&=\sum_{i=1}^n u(\theta_g; \mathbf{y_i})
\frac{c}{\sqrt{n}}+\sum_{i=1}^n \nabla u(\theta_g;\mathbf{y_i})
\frac{c^2}{2n}+ O_p(n^{-1/2})\\ \nonumber 
& \rightarrow^d
N\left(\frac{c^2}{2} \underbrace{\mathbf{E}(\nabla u(\theta_g; \mathbf{Y})
)}_{a=-H(\theta_g)},c^2\underbrace{var(u(\theta_g; \mathbf{Y}))}_{b=J(\theta_g)}
\right) \mbox{  } \textit{(from Eq. \eqref{eq:score}and \eqref{eq:hessian})} \\
&\rightarrow N\left(-\frac{c^2}{2}a,c^2b\right)
\end{align}
 $O_p(n^{-1/2})$ can be justified since the log composite likelihood is a finite sum of genuine  log likelihoods, which are of the same order.
 Note that Eq.\eqref{eq:bump_function1} does not generate the bump function since the mean is not the negative half of the variance in the asymptotic normal distribution.  
In order to obtain the bump function, we can adjust the ratio of composite likelihoods by raising it to the power (a/b);

\begin{align}
\frac{a}{b}\log \frac{CL(\theta;y)}{CL(\theta_g;y)} &
\rightarrow N \left( -\frac{a^2c^2}{2b},\frac{a^2c^2}{b} \right) \\ \nonumber
\end{align}

$$\therefore \lim_{n \rightarrow \infty} P\left( \left(\frac{CL(\theta;y)}{CL(\theta_g;y)}\right)^{a/b} \geq k\right)=\Phi \left(-\frac{c^*}{2}-\frac{\log k}{c^*} \right)  $$
where $c^*=\frac{ac}{\sqrt{b}}$\\
We can estimate $a/b$ through consistent estimates of $J(\theta)$ and $H(\theta)$. Let $\hat{\theta}_{CL}$ be the maximum likelihood estimator of $\theta$, which is a consistent estimator of $\theta_g$ (Xu, 2012) %\citep{xu:2012}, 
then
%$$\hat{H}(\theta_g)=\frac{1}{n}\sum_{i=1}^n \nabla_{\theta} u(\hat{\theta}_{CL};y_i) $$
$$\hat{a}=\frac{1}{n}\sum_{i=1}^n \nabla_{\theta} u(\hat{\theta}_{CL};y_i) $$
$$\hat{b}=\frac{1}{n}\sum_{i=1}^n  u^2(\hat{\theta}_{CL};y_i) $$ 
\end{enumerate}

\subsection{Proof of Theorem 2}
\begin{enumerate}[label=(\alph*)] 
\item We want to show,
\begin{eqnarray}\label{eq:appendix_profile1} 
\frac{CL_p(\psi_g,y)}{CL_p(\psi,y)} \to^p \infty \quad n \to \infty
\end{eqnarray}
$CL_p(\psi_g;y)=CL(\psi_g,\hat{\lambda}(\psi_g);y)=\sup_\lambda CL(\psi_g,\lambda;y) \geq CL(\psi_g,\lambda;y) \quad \forall \lambda$ thus  true for $\lambda_g$.  

Then it would be enough to show $\frac{CL(\psi_g,\lambda_g;y)}{CL_p(\psi;y)} \rightarrow^p \infty$ since $\frac{CL_p(\psi_g;y)}{CL_p(\psi;y)} \geq  \frac{CL(\psi_g,\lambda_g;y)}{CL_p(\psi;y)}$. This will imply that Eq.\eqref{eq:appendix_profile1} holds.

In \citet*{severini:2000} on page 127, it was shown that the difference between a profile log-likelihood function from a genuine log likelihood function is of order $O_p(1)$, i.e. $l_p(\psi;y)=l(\psi,\lambda(\psi);y)+O_p(1)$, here $l(\psi,\lambda(\psi);y)$ refers to a genuine log likelihood function as it can be obtained  from an actual model for the data using a Taylor expansion $l_p(\psi;y)=l(\psi,\hat{\lambda}(\psi);y)$ about $l(\psi,\lambda(\psi);y)$. Following a similar Taylor expansion for the composite likelihood, we get;
\begin{align*}
cl(\psi,\hat{\lambda}(\psi))&=cl(\psi,\lambda(\psi))+ cl_{\lambda}(\psi,\lambda(\psi))^T (\hat{\lambda}(\psi)-\lambda(\psi)) \\
& \quad + \frac{1}{2}  (\hat{\lambda}(\psi)-\lambda(\psi))^T  cl_{\lambda\lambda}(\psi,\lambda(\psi)) (\hat{\lambda}(\psi)-\lambda(\psi)) + ...
\end{align*}
Then if $\hat{\lambda}(\psi)=\lambda(\psi)+O_p(n^{1/2})$ is true then we conclude that $cl_p(\psi;y)=cl(\psi,\lambda(\psi))+O_p(1)$. Since $cl(\psi,\lambda(\psi))$ is a finite sum of genuine log-likelihood functions, under regularity condition on genuine likelihood functions, $cl_\lambda(\psi,\lambda(\psi))=O_p(\sqrt{n})$ and $cl_{\lambda\lambda}(\psi,\lambda(\psi))=O_p(n)$. 

\underline{Why is $\hat{\lambda}(\psi)=\lambda(\psi)+O_p(n^{1/2})$ true?}

Remember $\theta_g=(\psi_g,\lambda_g)$ is the value of the parameter that minimizes the K-L divergence between the assumed model $f$ and the true model $g$. In the profile composite likelihood $CL_p(\psi)=CL(\psi,\hat{\lambda}(\psi))=\sup_\lambda CL(\psi,\lambda)$, $\hat{\lambda}_\psi$ is the maximum likelihood estimate of $\lambda$ for a fixed $\psi$.  In general $\hat{\lambda}(\psi)$ is not a consistent estimator of $\lambda_g$ unless $\psi$ is fixed at the `true' value, $\psi_g$. Note that:

\begin{equation*}\frac{1}{n}cl(\psi,\lambda) - \frac{1}{n}E_g[cl(\psi,\lambda) ] \xrightarrow{p} 0.\end{equation*}

Following the arguments in \citet*{severini:2000}, section 4.2.1, %under some regularity conditions, 
the maximizer of $cl(\psi,\lambda)/n$ should converge in probability to the maximizer of  $E_g[cl(\psi,\lambda) ]/n$, which is $(\psi_g,\lambda_g)$. It was shown in \citet*{xu:2012} that the maximum composite likelihood estimator, $\hat{\theta}_{CL}$, converges almost surely to $\theta_g$ where $\theta_g$ is the parameter that minimizes the Kullback-Leibler divergence between the working model $f$ and the true model $g$ (Eq. \eqref{eq:KL_divergence}). Here, we treat $\psi$ fixed, then $\lambda(\psi)$ becomes the only parameter and $\hat{\lambda}(\psi)$ is the MLE of $\lambda(\psi)$ for a fixed $\psi$. Note that when $\psi=\psi_g$, $\lambda(\psi_g)=\lambda_g$. By following the regular arguments about the composite MLEs in \citet*{xu:2012}, it can be shown that $\hat{\lambda}(\psi) \to \lambda(\psi) $ as $n\to 0$, where $\lambda(\psi)$ is the value of $\lambda$ that maximizes $n^{-1}E_g[l(\psi,\lambda) ]   $ when $\psi$ is fixed. The asymptotic distribution of $\sqrt{n} \left(  \hat{\lambda}(\psi)-\lambda(\psi)\right)$ is derived in Eq.\eqref{eq:profile_MLE_derive} and Eq.\eqref{eq:restricted_mle}. Thus $\hat{\lambda}(\psi)=\lambda(\psi)+O_p(n^{-1/2})$ and $cl_p(\psi)=cl(\psi,\lambda(\psi))+O_p(1)$. \qed

%\begin{framed}
%\textbf{Reminder:} If $X \xrightarrow{p} a$ then $X_n=O_p(1)$ (bounded in probability). $\forall$ $\epsilon>0$, there exists a $ \gamma(\epsilon) < \infty $ such that $P( | X_n| \geq \gamma(\epsilon)) < \epsilon$.
%\end{framed}
%\begin{framed}
%\textit{In Severini(2000), page 88: The higher-order derivatives are, in general, of order $O_p(n)$ and are equal to a constant of order $O(n)$ plus a term of order $O_p(\sqrt{n})$. For instance;
%$$l_{\theta\theta}(\theta)=-i(\theta)+O_p(\sqrt{n})$$ where $-i(\theta)=E[l_{\theta\theta}(\theta);\theta] $\\
%Similar results hold for products of log-likelihood derivatives. For instance, $l_\theta(\theta)l_{\theta\theta}(\theta)$ is of order $O_p(n)$ and may be written 
%$$l_\theta(\theta)l_{\theta\theta}(\theta)=E[ l_\theta(\theta)l_{\theta\theta}(\theta);\theta ]+O_p(\sqrt{n})  $$
%}
%\end{framed}
It follows that:
\begin{align} \label{logratio}
(1/n) \log \frac{CL(\psi_g,\lambda_g)}{CL_p(\psi)} =(1/n) ( cl(\psi_g,\lambda_g) - cl(\psi,\lambda(\psi)) )+ (1/n)O_p(1)
\end{align}
where $cl(\theta)=\log CL(\theta)$. 

\begin{align*}
\log \left(\frac{\Pi_{i=1}^n CL(\psi_g,\lambda_g; y_i)} {\Pi_{i=1}^n CL(\psi,\lambda(\psi); y_i)} \right)^{1/n}&= \frac{1}{n} \left( \sum_{i=1}^n cl(\psi_g,\lambda_g;y_i)   - \sum_{i=1}^n cl (\psi,\lambda(\psi)  ;y_i)  \right) + (1/n)O_p(1) \\
 &=\frac{1}{n} \left( \sum_{k=1}^K w_k \left\{ \sum_{i=1}^n  \log(f(y_i \in A_k;\psi_g,\lambda_g))   \right. \right.  \\
 & \quad -  \left. \left.   \sum_{i=1}^n \log(f(y_i \in A_k; \psi,\lambda(\psi)  )) \right\}        \right) + (1/n)O_p(1) \\ 
 &\xrightarrow{p}  \sum_{k=1}^K w_k \left\{ \mathbf{E}_{g} \left[ \log(f(Y \in A_k;\psi_g,\lambda_g ))- \log(f(Y\in A_k; \psi,\lambda(\psi)))\right] \right\} \\ 
 & > 0
\end{align*}
since $w_k$'s are positive and $\theta_g=(\psi_g,\lambda_g)$ minimizes the Kullback-Leibler divergence in \eqref{eq:KL_divergence}.

Let $R_n= \Pi_{i=1}^n CL(\psi_g,\lambda_g; y_i)/ \Pi_{i=1}^n CL(\psi,\lambda(\psi); y_i) $. We get $1/n\log R_n \rightarrow c > 0$, where $c$ is a finite positive number, then $R_n=  \Pi_{i=1}^n CL(\psi_g,\lambda_g; y_i)/\Pi_{i=1}^n CL(\psi,\lambda(\psi); y_i)  \xrightarrow{p} \infty$.  \qed
%\textbf{Why \eqref{secondRn} is true:}  We got $1/n\log R_n \rightarrow c <0$, this implies $\log R_n \rightarrow -\infty $. Because if  $\log
%R_n$ converged to a finite number $c_1$, $c_1 \neq 0$ , then
%$1/n \log R_n$ would converge to $0$ as $n \rightarrow \infty$. But
%$1/n \log R_n \rightarrow c<0$. Therefore $\log R_n \rightarrow
%-\infty $. Hence $R_n \rightarrow 0$ with probability 1. \\

%Suppose we have $\theta=(\psi,\lambda)$ and let $cl_p(\hat{\theta}_\psi)$ denote the profile composite likelihood ratio statistics for $\psi$.
\item We want to show $ \lim_{n \rightarrow \infty} P_g\left(\frac{CL_p(\psi)}{CL_p(\psi_g)} \geq k\right)=\Phi \left(-\frac{c}{2}-\frac{\log k}{c} \right) $,
where $c$ is proportional to the distance between $\psi$ and $\psi_g$. %For now we assume both $\psi$ and $\lambda$ are scaler for convenience. This was said in the body of the thesis.

Following the proof of \citet*{royall:2000} for profile likelihoods:
Let $cl_p(\psi) = \log CL_p(\psi)$. For $\psi=\psi_g+c/\sqrt{n}$,
\begin{eqnarray} \label{eq:main_eq}
cl_p(\psi)-cl_p(\psi_g) = cA_n +(c^2/2)B_n + R_n
\end{eqnarray}
where $A_n=\frac{1}{\sqrt{n}} \left.\frac{ d cl_p(\psi)}{d\psi}\right |_{(\psi_g)} $, $B_n=\frac{1}{n} \left.\frac{ d^2 cl_p(\psi)}{d\psi^2}\right |_{(\psi_g)}$ and $R_n=O_p(n^{-1/2})$

%{ I am not sure how to argue with this, since these are  profile log - likelihoods; Regularity condition 1 holds for regular log likelihoods (or finite sum of regular log-likelihoods) how do we argue that this order is correct ?}

%\textcolor{red}{partial derivative imi yoksa normal derivative mi?}
 
\begin{eqnarray} \label{eq:An}
A_n&=&\frac{1}{\sqrt{n}} \left.\frac{ d cl_p(\psi)}{d\psi}\right |_{(\psi_g)} =\frac{1}{\sqrt{n}} \left.\frac{ d cl(\psi,\hat{\lambda}(\psi))}{d\psi}\right |_{(\psi_g)} 
\end{eqnarray}

\begin{eqnarray}\label{eq:An_derivative}
 \left.\frac{ d cl(\psi,\hat{\lambda}(\psi))}{d\psi}\right |_{(\psi_g)} &=& \left.\frac{\partial cl(\psi,\lambda)}{\partial\psi}\right |_{(\psi_g,\hat{\lambda}(\psi_g))} + \underbrace{\left.\frac{\partial cl(\psi,\lambda)}{\partial\lambda}\right |_{(\psi_g,\hat{\lambda}(\psi_g))} }_{=0} \left. \frac{d \hat{\lambda}(\psi)}{d\psi} \right |_{(\psi_g)} \nonumber \\
  &=&  \left.\frac{\partial cl(\psi,\lambda)}{\partial\psi}\right |_{(\psi_g,\hat{\lambda}(\psi_g))}  
\end{eqnarray}
 
We make a Taylor expansion of  $\left.\frac{\partial cl(\psi,\lambda)}{\partial\psi}\right |_{(\psi_g,\hat{\lambda}(\psi_g))}$  about $\left.\frac{\partial cl(\psi,\lambda)}{\partial\psi}\right |_{(\psi_g,\lambda_g)}  $
\begin{eqnarray}\label{eq:An_Taylor} 
 \left.\frac{\partial cl(\psi,\lambda)}{\partial\psi}\right |_{(\psi_g,\hat{\lambda}(\psi_g))} &=& \left.\frac{\partial cl(\psi,\lambda)}{\partial\psi}\right |_{(\psi_g,\lambda_g)} + \left.\frac{\partial^2 cl(\psi,\lambda)}{\partial\psi\partial \lambda}\right |_{(\psi_g,\lambda_g)} ( \hat{\lambda}(\psi_g) - \lambda_g )  \nonumber   \\
 & & \quad \quad + R^*_n 
\end{eqnarray}
We observe that $\lambda(\psi_g)=\lambda_g$ and that $cl(\psi,\lambda(\psi))$ is a finite sum of genuine log-likelihood functions. Then $R^*_n$ in Eq.\eqref{eq:An_Taylor} is of $O_p(1)$ since the higher order derivatives of log-likelihood function are of order $O_p(n)$ and $( \hat{\lambda}(\psi_g) - \lambda_g )= O_p(n^{-1/2})$ due to  Eq.\eqref{eq:restricted_mle}.
%%% JULY 15, 2014   $R_n^* =   \left.\frac{\partial^3 cl(\psi,\lambda)}{\partial\psi\partial \lambda}\right |_{(\psi_g,t)} ( \hat{\lambda}(\psi_g) - \lambda_g )^2/2 $ where $| \lambda_g-t| \leq \delta$.
%Bu alttaki satirlar Severini nin regularity cond. gore yaptigim. yukardaki argumant ok ama A6 ve consistency of hat{lambda} yi kullanmadan, yani bor sonraki sayfadaki hat(lambda)-lambda nin distribution nini bulmada kullandigimiz methodu kullanmadim, muhtelen o da olur ama beni supheye dusuren su idi. hani 3. turevinde "t" yi koyuyoruz ya, burda sanki 2. turevinde koymus gibi olacagiz, ilk turek psi ya gore, 2. ve 3. turev lambda ya gore, o yuzden lambda ya gore sadece 2 kere turev alinmis A6 daki sartlari saglar mi emin olamadim.
%This follows from Regularity condition 1. 
%$\frac{sup ||R_n||}{n(\hat{\lambda}(\psi_g) -\lambda_g)^2 }=O_p(1)$, then $R_n=nO_p(n^{-1/2})^2 O_p(1)=O_p(n^{-1/2})$.\qed 
%\textcolor{red}{ $R_n=O_p(1)$.  $(\hat{\lambda}(\psi_g) -\lambda(\psi_g))=O_p(n^{-1/2})$}
%\textcolor{red}{$R_n=O_p(1) WHY?$ (from Nancy's notes)}
%\textcolor{red}{Also Severini R2 $\frac{||R_n||}{n(\hat{\lambda}(\psi_g) -\lambda_g)^2 }=O_p(1)$, then $R_n=nO_p(n^{-1/2})^2 O_p(1)$ and $R_n=O_p(1)$}

Then \eqref{eq:An} becomes:
\begin{align}\label{eq:An_last} 
\frac{1}{\sqrt{n}} \left.\frac{ d cl_p(\psi)}{d\psi}\right |_{(\psi_g)}&=& \underbrace{ \frac{1}{\sqrt{n}}\left.\frac{\partial cl(\psi,\lambda)}{\partial\psi}\right |_{(\psi_g,\lambda_g)}}_{(I)} + \underbrace{\frac{1}{n}{  \left.\frac{\partial^2 cl(\psi,\lambda)}{\partial\psi\partial \lambda}\right |_{(\psi_g,\lambda_g)}}}_{(II)}\underbrace{ \sqrt{n}( \hat{\lambda}(\psi_g) - \lambda_g ) }_{(III)}
\end{align}
$(I)$ and $(II)$ in Eq.\eqref{eq:An_last} become;
\begin{eqnarray*}
\frac{1}{\sqrt{n}}\left.\frac{\partial cl(\psi,\lambda)}{\partial\psi}\right |_{(\psi_g,\lambda_g)} &\to& N(0,J_{\psi\psi}(\psi_g,\lambda_g)) \quad \quad \quad \text{(by CLT)}\\
\frac{1}{n}{  \left.\frac{\partial^2 cl(\psi,\lambda)}{\partial\psi\partial \lambda}\right |_{(\psi_g,\lambda_g)}} &\to&E_g(  \left.\frac{\partial^2 cl(\psi,\lambda)}{\partial\psi\partial \lambda}\right |_{(\psi_g,\lambda_g)} ) \quad \quad \quad \text{(by LLN)}\\
&=&-H_{\psi \lambda}(\psi_g,\lambda_g)
\end{eqnarray*}
What about $(III)$ in Eq.\eqref{eq:An_last}?

Note that $\hat{\lambda}(\psi_g)$ is the solution to $\left.\frac{d cl(\psi_g,\lambda)}{d \lambda } \right |_{\hat{\lambda}_(\psi_g)}=0 $. Then a Taylor expansion of   $\left.\frac{d cl(\psi_g,\lambda)}{d \lambda } \right |_{\hat{\lambda}_(\psi_g)}$  about  $\left.\frac{d cl(\psi_g,\lambda)}{d \lambda } \right |_{\lambda_g}  $ gives,

\begin{align}\label{eq:profile_MLE_derive}
\left.\frac{d cl(\psi_g,\lambda)}{d \lambda } \right |_{\hat{\lambda}_(\psi_g)}=0 &=& \left.\frac{d cl(\psi_g,\lambda)}{d \lambda } \right |_{\lambda_g} +\left.\frac{d^2 cl(\psi_g,\lambda)}{d \lambda^2 } \right |_{\lambda_g} (\hat{\lambda}(\psi_g)-\lambda_g) +R_n^{**}.
\end{align}
Dividing both sides by $\sqrt{n}$, we get
%\textcolor{red}{$\frac{sup ||R_n||}{n(\hat{\lambda}(\psi_g) -\lambda_g)^2 }=O_p(1)$, then $R_n=nO_p(n^{-1/2})^2 O_p(1)$ and $R_n=O_p(1)$}\\
%\textcolor{red}{Severini page 127$\hat{\lambda}(\psi)=\lambda_\psi+O_p(n^{-1/2})$}
\begin{align}\label{eq:Rn_doublestar}
\frac{1}{\sqrt{n}}\left.\frac{d cl(\psi_g,\lambda)}{d \lambda } \right |_{\hat{\lambda}_(\psi_g)}&=&\frac{1}{\sqrt{n}} \left.\frac{d cl(\psi_g,\lambda)}{d \lambda } \right |_{\lambda_g} +\frac{1}{n} \left.\frac{d^2 cl(\psi_g,\lambda)}{d \lambda^2 } \right |_{\lambda_g} \sqrt{n} (\hat{\lambda}(\psi_g)-\lambda_g) +\frac{R_n^{**}}{\sqrt{n}}.
\end{align}
In \eqref{eq:Rn_doublestar}, $\frac{R_n^{**}}{\sqrt{n}}= \frac{1}{n} \left.\frac{d^3 cl(\psi_g,\lambda)}{d \lambda^3 } \right |_{t} \sqrt{n} (\hat{\lambda}(\psi_g)-\lambda_g)^2$ where $|\lambda_g-t| \leq \delta$.  Then by $A6$ and $(\hat{\lambda}(\psi_g)-\lambda_g) \to^p 0$, the following argument holds \citep[chap.5]{knight:2000},
\begin{align}\label{eq:distribution_MLE}
(\hat{\lambda}(\psi_g)-\lambda_g) \frac{1}{n} \left.\frac{d^3 cl(\psi_g,\lambda)}{d \lambda^3 } \right |_{t} &\to^p 0.
\end{align}
Then (III) in Eq.\eqref{eq:An_last} becomes 
\begin{eqnarray} \label{eq:restricted_mle}
\sqrt{n}(\hat{\lambda}(\psi_g)-\lambda_g) &\doteq & -\frac{ \frac{1}{\sqrt{n}} \left.\frac{d cl(\psi_g,\lambda)}{d \lambda } \right |_{\lambda_g}  }{\frac{1}{n} \left.\frac{d^2 cl(\psi_g,\lambda)}{d \lambda^2 } \right |_{\lambda_g}  } \nonumber \\ 
&\overset{d}{\to}& \frac{N(0,J_{\lambda \lambda}(\psi_g,\lambda_g))}{H_{\lambda \lambda}(\psi_g,\lambda_g)}  
\end{eqnarray}

Substituting \eqref{eq:restricted_mle} in \eqref{eq:An_last}

\begin{eqnarray*}
A_n=\frac{1}{\sqrt{n}} \left.\frac{ d cl_p(\psi)}{d\psi}\right |_{(\psi_g)} &\overset{d}{\to}& z_1  -H_{\psi \lambda}(\psi_g,\lambda_g)  \frac{z_2}{H_{\lambda \lambda}(\psi_g,\lambda_g)}   \\
\end{eqnarray*}

where $$\left( \begin{array}{c} z_1 \\ z_2 \end{array} \right) \overset{d}{\to} N\left(     \left[ \begin{array}{c} 0 \\ 0 \end{array} \right],  \left[\begin{array}{cc}
J_{\psi \psi}(\psi_g,\lambda_g) & J_{\psi \lambda}(\psi_g,\lambda_g) \\
J_{\lambda \psi}(\psi_g,\lambda_g) & J_{\lambda \lambda} (\psi_g,\lambda_g)
\end{array}
\right]         \right) $$

Then 
%\begin{framed}
\begin{align}\label{eq:An_last_dist}
A_n& \overset{d} {\to}  N\left(0,  [ J_{\psi \psi}(\psi_g,\lambda_g) + \left[\frac{H_{\psi \lambda}(\psi_g,\lambda_g)}{H_{\lambda \lambda}(\psi_g,\lambda_g)} \right]^2J_{\lambda \lambda}(\psi_g,\lambda_g) - 2 \frac{H_{\psi \lambda}(\psi_g,\lambda_g)}{H_{\lambda \lambda}(\psi_g,\lambda_g)} J_{\psi \lambda}(\psi_g,\lambda_g)]\right) 
\end{align}
%\end{framed}
What about $B_n$ in \eqref{eq:main_eq}?
\begin{eqnarray} \label{eq:Bn}
B_n&=&\frac{1}{n} \left.\frac{ d^2 cl_p(\psi)}{d\psi^2}\right |_{(\psi_g)} 
\end{eqnarray}
where
\begin{eqnarray*}
\left.\frac{ d^2 cl_p(\psi)}{d\psi^2}\right |_{(\psi_g)} &=& \left. \frac{ d}{d \psi} \left[\frac{ d cl_p(\psi)}{d\psi}\right] \right |_{(\psi_g)} \\
&=& \frac{\partial}{\partial \psi} \left. \left[  \frac{\partial cl(\psi,\lambda)}{\partial\psi} + \frac{\partial cl(\psi,\lambda)}{\partial\lambda} \frac{d \hat{\lambda}(\psi)}{d\psi} \right] \right |_{(\psi_g,\hat{\lambda}(\psi_g))} \\
  &=&  \left.\frac{\partial^2l cl(\psi,\lambda)}{\partial\psi^2}\right |_{(\psi_g,\hat{\lambda}(\psi_g))} +  \left.\frac{\partial^2 cl(\psi,\lambda)}{\partial\psi\partial \lambda}\right |_{(\psi_g,\hat{\lambda}(\psi_g))}  \left.\frac{d \hat{\lambda}(\psi) }{d\psi}\right |_{(\psi_g)} \\ 
 & & +  \left \{  \left.\frac{\partial^2 cl(\psi,\lambda)}{\partial\psi\partial \lambda}\right |_{(\psi_g,\hat{\lambda}(\psi_g))} + \left.\frac{\partial^2 cl(\psi,\lambda)}{\partial\lambda^2}\right |_{(\psi_g,\hat{\lambda}(\psi_g))}    \frac{d \hat{\lambda}(\psi) }{d\psi}     \right \}  \left.\frac{d \hat{\lambda}(\psi) }{d\psi}\right |_{(\psi_g)}   \\
  & & +   \left.\frac{d^2 \hat{\lambda}(\psi) }{d\psi^2}\right |_{(\psi_g)}  \underbrace{\left.  \frac{\partial cl(\psi,\lambda)}{\partial\lambda}   \right |_{(\psi_g, \hat{\lambda}(\psi_g))  } }_{=0} \\
&=& \left.\frac{\partial^2l cl(\psi,\lambda)}{\partial\psi^2}\right |_{(\psi_g,\hat{\lambda}(\psi_g))} + 2  \left.\frac{\partial^2 cl(\psi,\lambda)}{\partial\psi\partial \lambda}\right |_{(\psi_g,\hat{\lambda}(\psi_g))}  \left.\frac{d \hat{\lambda}(\psi) }{d\psi}\right |_{(\psi_g)} \\ 
&& +  \left.\frac{\partial^2 cl(\psi,\lambda)}{\partial\lambda^2}\right |_{(\psi_g,\hat{\lambda}(\psi_g))}   \left( \left.\frac{d \hat{\lambda}(\psi) }{d\psi}\right |_{(\psi_g)} \right)^2 \\ 
\end{eqnarray*} \qed

Why is $R_n$ of $O_p(n^{-1/2})$ in Eq.\eqref{eq:main_eq}?
$R_n=\frac{1}{6n^{3/2}} \left.\frac{ d^3 cl_p(\psi)}{d\psi^3}\right |_{t}$ where $|\psi_g-t|\leq \delta$.
\begin{align*}
\left.\frac{ d^3 cl_p(\psi)}{d\psi^3}\right |_{t} &= \frac{ d}{d \psi} \left[ \left.\frac{\partial^2l cl(\psi,\lambda)}{\partial\psi^2}\right |_{(t,\hat{\lambda}(t))} + 2  \left.\frac{\partial^2 cl(\psi,\lambda)}{\partial\psi\partial \lambda}\right |_{(t,\hat{\lambda}(t))}  \left.\frac{d \hat{\lambda}(\psi) }{d\psi}\right |_{t} \right. \\ 
& \quad \quad \left. +  \left.\frac{\partial^2 cl(\psi,\lambda)}{\partial\lambda^2}\right |_{(t,\hat{\lambda}(t))}   \left( \left.\frac{d \hat{\lambda}(\psi) }{d\psi}\right |_{t} \right)^2 \right] \\ 
&=\underbrace{ \frac{ d}{d \psi} \left[ \left.\frac{\partial^2l cl(\psi,\lambda)}{\partial\psi^2}\right |_{(t,\hat{\lambda}(t))} \right]  }_{A1}+ \underbrace{ \frac{ d}{d \psi}  \left[  2  \left.\frac{\partial^2 cl(\psi,\lambda)}{\partial\psi\partial \lambda}\right |_{(t,\hat{\lambda}(t))}  \left.\frac{d \hat{\lambda}(\psi) }{d\psi}\right |_{t} \right] }_{A2} \\
& \quad \quad +  \underbrace{\frac{ d}{d \psi}\left[  \left.\frac{\partial^2 cl(\psi,\lambda)}{\partial\lambda^2}\right |_{(t,\hat{\lambda}(t))}   \left( \left.\frac{d \hat{\lambda}(\psi) }{d\psi}\right |_{t} \right)^2 \right] }_{A3} 
\end{align*}

\begin{align*}
A1 &=   \left.\frac{\partial^3 cl(\psi,\lambda)}{\partial\psi^3}\right |_{(t,\hat{\lambda}(t))} +  \left.\frac{\partial^3 cl(\psi,\lambda)}{\partial\psi^2\partial \lambda}\right |_{(t,\hat{\lambda}(t))}  \left.\frac{d \hat{\lambda}(\psi) }{d\psi}\right |_{t}   \\
A2 &=  2 \left[   \left\{  \left.\frac{\partial^3 cl(\psi,\lambda)}{\partial\psi^2\partial \lambda}\right |_{(t,\hat{\lambda}(t))} +  \left.\frac{\partial^3 cl(\psi,\lambda)}{\partial\psi\partial \lambda^2}\right |_{(t,\hat{\lambda}(t))}  \left.\frac{d \hat{\lambda}(\psi) }{d\psi}\right |_{t}  \right\}  \left.\frac{d \hat{\lambda}(\psi) }{d\psi}\right |_{t}   \right. \\
& \quad \quad \quad \quad + \left.      \left.\frac{\partial^2 cl(\psi,\lambda)}{\partial\psi\partial \lambda}\right |_{(t,\hat{\lambda}(t))}  \left.\frac{d^2 \hat{\lambda}(\psi) }{d\psi^2}\right |_{t}         \right] \\
A3 &=   \left[   \left.\frac{\partial^3 cl(\psi,\lambda)}{\partial\psi\partial \lambda^2}\right |_{(t,\hat{\lambda}(t))} +  \left.\frac{\partial^3 cl(\psi,\lambda)}{\partial \lambda^3}\right |_{(t,\hat{\lambda}(t))}  \left.\frac{d \hat{\lambda}(\psi) }{d\psi}\right |_{(t)}   \right] \left( \left.\frac{d \hat{\lambda}(\psi) }{d\psi}\right |_{t} \right)^2  \\
& \quad \quad \quad \quad +    2  \left.\frac{\partial^2 cl(\psi,\lambda)}{\partial\psi\partial \lambda}\right |_{(t,\hat{\lambda}(t))}  \left.\frac{d \hat{\lambda}(\psi) }{d\psi}\right |_{t} \left.\frac{d^2 \hat{\lambda}(\psi) }{d\psi^2}\right |_{t}   \\
\end{align*}
From Lemma \ref{lemma:lemma2} below, $\left. d \hat{\lambda}(\psi)/d\psi \right |_{t}=O_p(1)$. By taking the second derivative of $d \hat{\lambda}(\psi)/d\psi $ with respect to $\psi$, it is seen that $\left. d^2 \hat{\lambda}(\psi)/d^2\psi \right |_{t}=O_p(1)$.
%July 14, 2014
Also, following the arguments presented in Lemma \ref{lemma:lemma1}, we observe that the second and higher order derivatives of composite log-likelihood functions in $A_1$, $A_2$ and $A_3$ are of $O_p(n)$, since the mean of the log likelihood derivatives are of order $O(n)$ (they are $O(1)$ for one observation) and the log composite likelihood is a finite sum of log likelihoods, e.g. the first term in $A_1$ is $O_p(n)$ since
$$\frac{1}{n} \left.\frac{\partial^3 cl(\psi,\lambda)}{\partial\psi^3}\right |_{(t,\hat{\lambda}(t))} \to E_g \left(  \left. \frac{\partial^3 cl(\psi,\lambda)}{\partial\psi^3} \right |_{(t,\lambda(t)} \right).$$

%bu onceki aciklama, simdiki aciklama yukarda
% Since the second and higher order derivatives of the log-likelihood function are of order $O_p(n)$ and since the log composite likelihood is a finite sum of log likelihoods, they will be of the same order. Thus we conclude that $A1, A2$ and $A3$ are of $O_p(n)$.\\
$\therefore R_n=O_p(n^{-1/2})$.\qed

%July 14, 2014. From Lemma \ref{lemma:lemma2} below, it can be seen that $\left. d \hat{\lambda}(\psi)/d\psi \right |_{t}=O_p(1)$ when $\psi_g=t$. yani Lemma 2 ve Lemma 3 hala gecerli, psi_g yerine t koyunca. Sorun su simdi... A1, A2 ve A3 deki likelihoodlar, O_p(1) mi? MLE plug in edilmis bile mi olsa?
\begin{lemma} \citep{royall:2000} \label{lemma:lemma1} \normalfont
\begin{align*}
\frac{1}{n}  \left.\frac{\partial^2 cl(\psi,\lambda)}{\partial\psi^2}\right |_{(\psi_g,\hat{\lambda}(\psi_g))} \to -H_{\psi \psi}(\psi_g,\lambda_g)
\end{align*}
\begin{align*}
 \frac{1}{n}  \left.\frac{\partial^2 cl(\psi,\lambda)}{\partial\psi \partial \lambda}\right |_{(\psi_g,\hat{\lambda}(\psi_g))} \to -H_{\psi \lambda} (\psi_g,\lambda_g)
\end{align*}
This follows from the Law of Large Numbers, since $\hat{\lambda}(\psi)$ is the MLE in the one dimensional model with a fixed $\psi$.
\end{lemma}
%($\hat{\lambda}(\psi)=\lambda_{\psi}+O_p(n^{-1/2})$ (Severini, page 127) ) and \\
 %\textcolor{red}{ when $\psi=\psi_g$, $\hat{\lambda}(\psi)$ converges to $\lambda(\psi_g)=\lambda_g$??? }\\
 %\textcolor{red}{Are these arguments enough ???}

\begin{lemma} \citep{royall:2000}  \label{lemma:lemma2} \normalfont
\begin{align}
\left. \frac{d \hat{\lambda}(\psi)}{d\psi}  \right |_{(\psi_g)} \to -\frac{H_{\psi \lambda}(\psi_g,\lambda_g) }{H_{\lambda \lambda}(\psi_g,\lambda_g) } 
\end{align}
\begin{proof}
Let  $g(\psi,\hat{\lambda}(\psi))=\left.  \frac{\partial cl(\psi,\lambda)}{\partial \lambda} \right |_{\hat{\lambda}(\psi)}  $. Since $g(\psi,\hat{\lambda}(\psi)) =0 \quad  \forall \psi $, $g(\psi,\hat{\lambda}(\psi))$ is a constant. Thus $dg/d\psi=0.$
\begin{eqnarray*}
\frac{dg}{d\psi}&=&\left. \frac{\partial g(\psi,\lambda)}{\partial \psi}   \right |_{\hat{\lambda}(\psi)}+\left. \frac{\partial g(\psi,\lambda)}{\partial \psi}  \right |_{\hat{\lambda}(\psi)} \frac{d\hat{\lambda}(\psi)}{d\psi}\\
0&=&\left. \frac{	\partial^2cl(\psi,\lambda)}{\partial \psi \partial \lambda}   \right |_{\hat{\lambda}(\psi)}+ \left. \frac{\partial^2 cl(\psi,\lambda)}{\partial \psi^2}  \right |_{\hat{\lambda}(\psi)} \frac{d\hat{\lambda}(\psi)}{d\psi}
\end{eqnarray*}
Thus $\dfrac{d\hat{\lambda}(\psi)}{d\psi}= - \dfrac{\left. \frac{\partial^2cl(\psi,\lambda)}{\partial \psi \partial \lambda}   \right |_{\hat{\lambda}(\psi)}}{\left. \frac{\partial^2 cl(\psi,\lambda)}{\partial \psi^2}  \right |_{\hat{\lambda}(\psi)}}$ The conclusion follows from Lemma \ref{lemma:lemma1}.
\end{proof}
\end{lemma}

Then 
\begin{eqnarray*} 
B_n \to -H_{\psi \psi} + H_{\psi \lambda} \frac{H_{\psi \lambda}}{H_{\psi \psi}}
\end{eqnarray*}
Then $\eqref{eq:main_eq}$ becomes;

\begin{eqnarray} \label{eq:main_eq_last}
cl_p(\theta_1)-cl_p(\theta_g) &\overset{d}{\to}& N(-\frac{c^2}{2} a, c^2b)
\end{eqnarray}

where 
\begin{eqnarray*}
a&=&H_{\psi \psi} - H_{\psi \lambda} \frac{H_{\psi \lambda}}{H_{\psi \psi}}=H^{\psi \psi}(\psi_g,\lambda_g)^{-1}\\
b&=&[ J_{\psi \psi}(\psi_g,\lambda_g) + \left[\frac{H_{\psi \lambda}(\psi_g,\lambda_g)}{H_{\lambda \lambda}(\psi_g,\lambda_g)} \right]^2J_{\lambda \lambda}(\psi_g,\lambda_g) - 2 \frac{H_{\psi \lambda}(\psi_g,\lambda_g)}{H_{\lambda \lambda}(\psi_g,\lambda_g)} J_{\psi \lambda}(\psi_g,\lambda_g)] \quad (\text{from  } \eqref{eq:An_last_dist})\\
&=& H^{\psi \psi}(\psi_g,\lambda_g)^{-1} G^{\psi \psi}(\psi_g,\lambda_g)  H^{\psi \psi}(\psi_g,\lambda_g)^{-1}
\end{eqnarray*}

We see that \eqref{eq:main_eq_last} does not produce a bump function (the mean is not the negative half of the variance). If we take $\left(\frac{cl_p(\hat{\theta}_{\psi_1})}{cl_p(\hat{\theta}_{\psi_g})}\right)^{a/b}$ then 
$$\left(\frac{cl_p(\psi_1)}{cl_p(\psi_g)}\right)^{a/b} \to N(-\frac{c^2}{2}\frac{a^2}{b},\frac{c^2a^2}{b}),$$ which results in a bump function. Then the probability of misleading evidence will be
$$P_g\left\{\left(\frac{cl_p(\psi_1)}{cl_p(\psi_g)}\right)^{a/b} \geq k \right\} \to \Phi \left \{ -\frac{(ca/b^{1/2})}{2}-\frac{\log (k)}{(ca/b^{1/2})}  \right \} \quad \quad \textit{(the bump function)}.$$
In Theorem 2, $c*=ca/b^{1/2}$.
\end{enumerate}

\section{Generating correlated binary data}\label{App:Appendix_generate}
The method in \citet{emrich:1991} for generating correlated binary data uses a discretised normal approach to  generate correlated binary variates with specified marginal probabilities and pairwise correlations.  Suppose we want to generate  a $k$-dimensional correlated binary vector, $\mathbf{Y}=(Y_1,...,Y_k)$ given  $\mathbf{X}=(X_1,...,X_k)$, such that $p_i=E[Y_i \mid x_i]$ for $i=1,...,k$ and $p_{ij}=E[Y_i Y_j \mid x_i,x_j ]=Pr(Y_i=1,Y_j=1 \mid x_i,x_j)$ for $i=1,...,k-1$ and $j=2,...k$.  There are different approaches to quantify the dependence between a pair of binary observations. One approach is to quantify the dependence using the correlation between $Y_i$ and $Y_j$, however, the correlation gets constrained depending on the marginal probabilities, $p_i$ and $p_j$ (Prentice, 1988). Here, we use the association via the odds ratio (Dale, 1986). It is the ratio of the odds $Y_i=1$ given that $Y_j=1$ and the odds of $Y_i=1$ given that $Y_j=0$, which is interpreted as the odds of concordant pairs to discordant pairs.
 
 \begin{eqnarray*} 
\psi_{ij} &=&   \frac{ Pr(Y_{i}=1 \mid Y_{j}=1, x_i,x_j) \big{/} Pr(Y_{i}=0 \mid Y_{j}=1, x_i,x_j)  }  { Pr(Y_{i}=1 \mid Y_{j}=0, x_i,x_j) \big{/}  Pr(Y_{i}=0 \mid Y_{j}=0, x_i,x_j) } \\
& =& \frac{  Pr(Y_{i}=1, Y_{j}=1\mid x_i,x_j) Pr(Y_{i}=0,Y_{j}=0 \mid x_i,x_j)  }{Pr(Y_{i}=1 , Y_{j}=0 \mid x_i,x_j)  Pr(Y_{i}=0 , Y_{j}=0 \mid x_i,x_j) } \\
&=&\dfrac{p_{ij}(1-p_i-p_j+p_{ij}) }{ (p_{i}-p_{ij})(p_j-p_{ij}}  \quad \quad \text{from Table \ref{tab:jointprob}}   \\
\end{eqnarray*}

\begin{table} [H]
	\caption{Different outcomes with probabilities of occurrence \citep*{cessie:1994}.}
    	\label{tab:jointprob}
\begin{center}
  \begin{tabular}{c  c  c  c c   c c}
    \hline
     && $Y_{j}=1$ && $Y_{j}=0 $  && \mbox{}   \\ \hline
    $Y_{i}=1$ && $p_{ij}  $ && $p_{i}-p_{ij} $ && $p_{i}$ \\ 
    $Y_{i}=0$ &&  $p_{j}-p_{ij}$ && $1-p_i-p_j+p_{ij}$  && $1- p_{i}$ \\  \hline
    \mbox{}  && $p_{j}$ && $1-p_{j} $  && 1 \\ \hline
    \hline
  \end{tabular}
\end{center}
\end{table}

Thus, the joint probability of $p_{ij}$ is written in terms of the marginal probabilities, $p_{i}$ and $p_{j}$ and the odds ratio, $\psi_{ij}$. (Plackett (1965).

%%%% PLACKETT REFERANCINI CHECK ET 

\begin{equation}
\label{eq:jointprobeq}
p_{ij}=\left\{
	\begin{array}{lr}
	\dfrac{1+(p_{i}+p_{j})(\psi_{ij}-1)-S(p_{i},p_{j},\psi_{ij})}{2(\psi_{ij}-1)} & \textit{if } \psi_{ij} \neq 1,\\
	p_{i}p_{j}  &  \textit{if } \psi_{ij}=1,
	\end{array}
	\right.	
\end{equation}
where
 \begin{align*}
 S(p_{i},p_{j},\psi_{ij})&=\sqrt{   \{   1+(p_i+p_j)(\psi_{ij}-1)  \}^2+4\psi_{ij}(1-\psi_{ij})p_ip_j     },
 \end{align*}
for $p_{i},p_{j} \in (0,1)$, $ \psi_{ij} \geq 0$. If $Y_{i}$ and $Y_{j}$ are independent then $\psi_{ij}=1$.
% We use  $\delta=\log \psi \in \mathcal{R}$ in the computations to remove the restriction on the parameter space of $\psi$.

%Thus, the pairwise likelihood in $\eqref{eq:cessie_lik}$  is a function of marginal probabilities, $p_{i1}$ and $p_{i2}$ (Eq.\eqref{eq:logistic}) and the parameter $\psi$ which characterizes the dependence between pairs in odds scale.

Also, the pairwise correlation, $ corr(Y_i,Y_j \mid x_{i},x_j)=\delta_{ij}$ is
\begin{align}\label{eq:y1y2_corr}
\delta_{ij}&=\frac{p_{ij}-p_ip_j}{\left(p_i(1-p_i)p_j(1-p_j)\right)^{1/2}},
\end{align}

Now, let $\mathbf{Z}=(Z_1,...,Z_k)$ be a standard multivariate random variable with mean $0$ and correlation matrix $\Sigma=(\rho_{ij})$ with $i=1,...,k-1$ and $j=2,...k$. Then set $Y_i=1$ if $Z_i \leq z(p_i)$ and set $Y_i=0$ otherwise for $i=1,...,k$, where $z(p_i)$ is the $p_i^{th}$ quantile of the standard normal distribution. This leads to
\begin{align}\label{eq:mean_inverse}
E[Y_i \mid x_i ]&=Pr(Y_i=1 \mid x_i)=Pr(Z_i \leq z(p_i))=p_i,
\end{align}
and
\begin{align*}
E[Y_i \: Y_j \mid x_{i}, x_j ]&=Pr(Y_i=1,Y_j=1 \mid x_{i},x_j)=Pr(Z_i \leq z(p_i),Z_j \leq z(p_j))=\Phi\left(z(p_i),z(p_j),\rho_{ij}\right),
\end{align*}
where 
\begin{align*}
\Phi(z(p_1),z(p_2),\rho_{ij})&=\int_{-\infty}^{z(p_1)}\int_{-\infty}^{z(p_2)} f(z_1,z_2,\rho)dz_1dz_2,
\end{align*}
and $f(z_1,z_2,\rho)$ is the probability density function of a standard bivariate normal random variable with mean $\mathbf{0}$ and correlation coefficient $\rho$.

Note from Eq.\eqref{eq:y1y2_corr} that  $E[Y_i Y_j \mid x_{i} x_{j} ]= p_ip_j+\delta_{ij}\left(p_i(1-p_i)p_j(1-p_j)\right)^{1/2}$.  Then,
\begin{align}\label{eq:solve_eq}
\Phi\left(z(p_i),z(p_j),\rho_{ij}\right)&=p_ip_j+\delta_{ij}\left(p_i(1-p_i)p_j(1-p_j)\right)^{1/2}.
\end{align}
To solve Eq.\eqref{eq:solve_eq}  for $\rho_{ij}$, \citet{emrich:1991} suggested using a bisection technique. This method does not ensure that the pairwise probabilities $p_{ij}$,  or the correlation matrix composed of binary correlations $(\delta_{ij})$ are valid. Below are the compatibility conditions that are needed to be checked \citep{leisch:1998};
\begin{enumerate}
\item $0 \leq p_i \leq 1$ for  $i=1,...,k$.
\item $\max(0,p_i+p_j-1) \leq p_{ij} \leq \min(p_i,p_j)$ for $i \neq j$.
\item $p_i+p_j+p_l-p_{ij} - p_{il} - p_{jl} \leq 1$ for $i \neq j, j \neq l, l \neq i $.
\end{enumerate}
These conditions are necessary in order to get a nonnegative joint mass function for $\mathbf{Y}$ \citep{emrich:1991}. 
%Even if these conditions are satisfied, the simulation method might fail to produce a multivariate binary distribution especially in high dimensions \citep{jin:2010}.

We define the bivariate joint probability, $p_{ij}$, in terms of the marginal probabilities $p_i$ and $p_j$, and an odds ratio $\psi_{ij}$ (Eq \eqref{eq:jointprobeq}). However, for this simulation method, we need to use pairwise correlations $\delta_{ij}$ instead of $\psi$. The relationship between $\delta_{ij}$, and $\psi_{ij}$ can be easily established when we plug Eq.\eqref{eq:jointprobeq} in Eq.\eqref{eq:y1y2_corr}. 

Below, we provide a step-by-step summary of the simulation algorithm for generating a $k$-dimensional binary vector:
\begin{enumerate}
\item Set $\beta_0$, $\beta_1$ and $\psi$ values. Determine the marginal probabilities from the logistic model and second order probabilities in Eq.\eqref{eq:jointprobeq}. Check the compatibility conditions. Calculate pairwise correlations $\delta_{ij}$ between each binary pair.
\item Calculate $z(p_i)$ from Eq.\eqref{eq:mean_inverse}.
\item Solve Eq.\eqref{eq:solve_eq} to obtain the elements of the correlation matrix $\Sigma=(\rho_{ij})$ for the multivariate normal distribution.
\item Generate a $k$-dimensional multivariate normal vector $\mathbf{Z}$ with mean $z(p_i)$ and correlation matrix $\Sigma=(\rho_{ij})$.
\item Set $y_i=1$ if $z_i\leq z(p_i)$, and $y_i=0$ otherwise, for $i=1,...,k$.
\end{enumerate} 

We check the compatibility conditions in Step 1 using the R package \textit{bindata} (\cite{leisch:2012}), and for Steps 2--5, we used the $R$ package \textit{mvtBinaryEP} (\cite{by:2011}) in the $R$ Statistical Software. This procedure generates a vector of binary variables with the desired properties, namely, $E(Y_i)=p_i$ and $corr(Y_i,Y_j \mid x_i x_j)=\delta_{ij}$. Other methods for generating correlated data can be found in \citet{jin:2010}.% for a broader literature review on generating correlated binary data.

%Note that there are 5 values assumed for the dependence parameter; $\psi_1$ which quantifies the dependence between siblings and $\psi_2$ which  quantifies the dependence between a parent and an offspring,  $\psi_3$ which quantifies the dependence between an aunt/uncle and a niece/nephew, $\psi_4$ which quantifies the dependence between the grandparent and the grandchild and $\psi_5$ which quantifies the dependence between cousins. This should be taken into account in Step 1. 

Another important point is that this method generates correlated binary data that satisfy the first and second order marginal distributions. There might be more than one joint distribution that generate the same lower dimensional marginal distributions, in which case the inference from a composite likelihood approach will be the same for that family of distributions \citep{varin:2011}.  This property of composite likelihood inference is viewed as being robust by many authors (\citet{xu:2012},  \citet{varin:2011}, \citet{jin:2010}).
%\textcolor{blue}{Instead of trying to specified the complicated high dimensional joint density... Ximing's thesis
% (bu cumle ayni alinda varin:2011 den sayfa 24 / robustness) --- The composite likelihoods requires assumptions only in the lower dimensional distributions, the inference from the composite likelihood approach will be the same for that family of distributions, which earns the composite likelihood inference robust.
% the comp lik has less restrictions.
\section{Finding the profile maximum likelihood estimates}\label{App:Appendix_profileMLE}
We follow the steps below to find the profile maximum likelihood estimates (MCLE) of the parameter of interest, for example, when the parameter of interest is $\beta_1$ and $\theta=(\beta_0,\beta_1)$.

\begin{enumerate}
\item Set a grid for $\beta_{1}$, i.e. $\{\beta_{11},\beta_{12},...,\beta_{1g}\}$.
\item For each $\beta_{1i}$, $i=1,...,g$, maximize the composite likelihood chosen  with respect to the nuisance parameters as a function of $\beta_{1i}$. Obtain the composite likelihood value for each $(\beta_{1i}, \hat{\beta_0}(\beta_{1i}))$. Note that ($\hat{\beta_0}(\beta_{1i})$ is the MCLE when $\beta_1$ is taken as fixed.  For this, we use the Newton Raphson algorithm which we coded using the $R$ Statistical Software (\citet{R:2015})
\item Find the profile MCLE, $\hat{\beta}_{1CL_p}$, which maximizes the composite likelihoods calculated in Step 2.
\end{enumerate}

\section{The composite likelihood constructed from pairwise margins}\label{App:Appendix_pairwise}

When the dependence parameter is also of interest, we need to construct the composite likelihood from pairwise (or higher order) likelihood components \citep{varin:2011}.
The composite likelihood constructed from pairwise likelihood components is,
\begin{align}\label{eq:pairwiselik}
CL_{pair}(\beta_0,\beta_1,\psi)&=\Pi_{i=1}^N  \left[ \Pi_{j=1}^{n_i-1} \Pi_{k=j+1}^{n_i}P(Y_{ij}=y_{ij},Y_{ik}=y_{ik} \mid \mathbf{x_i}) \right]^{1/(n_i-1)},
\end{align}  
where $\Pi_{j=1}^{n_i-1} \Pi_{k=j+1}^{n_i} P(Y_{ij}=y_{ij},Y_{ik}=y_{ik} \mid \mathbf{x_i})$ denotes the pairwise likelihood for the $i^{th}$ family (Table \ref{tab:jointprob} and Eq. \eqref{eq:jointprobeq}). The weight $1/(n_i-1)$ is used to weigh the contribution of each family according to its size when the parameter of interest is the marginal parameter, since each observation in a family of size $k$ presents in $k-1$ pairs \citep{zhao:2005}. 

When the parameter of interest is $\beta_1$,  we can determine the profile composite likelihood 
$CL_p(\beta_1)$$=\max_{\beta_0,\beta_1}\{CL_{pair}(\beta_0,\beta_1,\psi)\}$ and compute the profile composite likelihood estimate, $\hat{\beta_1}_{CL_{pair}}= \max_{\beta_1} \log CL_{p}(\beta_0,\beta_1,\psi)$. However, when we want to get inference about $\psi$, we use the composite likelihood in Eq.\eqref{eq:pairwiselik} without the weights for families. We do not need to use the weights since the dependence parameter $\psi$ appears the right amount of times in the pairwise likelihood function for a family of size $k$, where there are $k-1$ pairs.
\begin{align}\label{eq:pairwiselik_psi}
CL^{\psi}_{pair}(\beta_0,\beta_1,\psi)&=\Pi_{i=1}^N  \left[ \Pi_{j=1}^{n_i-1} \Pi_{k=j+1}^{n_i}P(Y_{ij}=y_{ij},Y_{ik}=y_{ik} \mid \mathbf{x_i}) \right]
\end{align}  
and calculate $CL_p(\psi)=\max_{\beta_0,\beta_1}\{CL^{\psi}_{pair}(\beta_0,\beta_1,\psi)\}$.

\section{More simulation results}\label{App:Appendix_sim}

In these simulations, we consider three different family structures, where $k$ is the family size.
\begin{enumerate}
\item \textit{Sibling study with $k=5$}: Data consist of only siblings, where the number of siblings is 5 in each family.  
\item \textit{Sibling study with $k \in \{2,3,4,5\}$}: Data consist of only siblings, where the number of siblings is 2, 3, 4 or 5 in each family. 
\item\textit{ Family study with $k=5$}: Data consist of nuclear families with 3 siblings, i.e., 2 parents and 3 offspring.
\end{enumerate}

In the Sibling study with $k=5$, we choose three different values for the dependence parameter $\psi$ to indicate weak dependence ($\psi=1.2$), moderate dependence ($\psi=3$) and strong dependence ($\psi=6$) within family members. Here we are interested in whether the inference about $\beta_1$ is affected by different strengths of dependence. For the other family structures, we only take into account $\psi=3$.  In Table \ref{table:sibling_k5}, for the Sibling study with $k=5$ siblings, we see that as sample size increases, both composite likelihood approaches provide consistent estimates for the true parameter value $\beta_1$. The results do not change whether the dependence between sibling pairs are weak or strong.

\begin{table}[h!] \small
\centering
\caption{Sibling study with $k=5$. The maximum profile composite likelihood estimates of $\beta_1$, using independent and pairwise likelihood methods, under a weak, moderate and strong dependence parameter ($\psi=1.2,3$ and $6$) and with different numbers of families ($n$) and $\beta_0=-1$, $\beta_1=2$. \label{table:sibling_k5}}
\vspace{0.1in}
\begin{tabular}{r   cc    c   c   c    cc    c c c  c  } 
&  &&    \multicolumn{8}{c}{$\hat{\beta_1}_{CL_p}$}    &      \\[1ex]  \hline
&  &&    \multicolumn{2}{c}{ $\psi=1.2$ }  &&  \multicolumn{2}{c}{ $\psi=3$ }  &&  \multicolumn{2}{c}{ $\psi=6$ }   &  \\[1ex]   \hline
$n$&  & &     independent    &   pairwise   &&  independent    &   pairwise &&  independent    &   pairwise &   \\[1ex]
\hline
\multirow{1}{*}{ $30$} 
 && &             2.048         &     2.048            &&      2.064    &    2.062     &&  2.096         &     2.090        &      \\  [0.5ex]

\multirow{1}{*}{ $100$} 
 && &             2.016        &     2.016            &&      2.023    &    2.023     &&  2.024         &     2.024        &      \\   [0.5ex]

\multirow{1}{*}{$300$} 
 && &             2.002         &     2.002            &&     2.006    &    2.006     &&  2.006        &     2.006      &      \\   [0.5ex]

\multirow{1}{*}{$500$} 
 && &             2.002         &     2.002            &&      2.003    &     2.003     &&  2.003        &     2.003      &      \\  [0.5ex]

\multirow{1}{*}{$1000$} 
 && &             2.000         &     2.000            &&      2.002    &    2.002     &&  2.002          &     2.002         &     \\  [0.5ex]
\end{tabular}
\end{table}

In Table \ref{table:sibling_kdiff}, for the Sibling study with $k \in \{2,3, 4, 5\}$, we also see that as sample size increases, both composite likelihood approaches provide consistent estimates for the true parameter value $\beta_1$.

\begin{table}[h!]\small
 \caption{Sibling study with $k \in \{2,3, 4, 5\}$. The maximum profile composite likelihood estimates of $\beta_1$, using independent and pairwise likelihood methods, under the moderate dependence parameter $\psi=3$ and with different number of families $n$ and $\beta_0=-1$, $\beta_1=2$. \label{table:sibling_kdiff}}
 \centering
 \begin{tabular}{ r c   ccc    c }
  &&     \multicolumn{3}{c}{$\hat{\beta_1}_{CL_p}$}  &  \\[1ex]  \hline
 n &&    & independent    &   pairwise    \\[1ex] \hline 
30  &&    & 2.106  & 2.102      \\ 
100  &&  &2.029  & 2.029      \\
300  &&   &2.008  & 2.008      \\ 
500 & &   &2.006  & 2.006       \\ 
1000 & & & 2.002  & 2.002        \\ 
\end{tabular}
\end{table} 

In Table \ref{table:family_k5}, for the Family study with $k=5$, we see that as sample size increases, the independent likelihood provides consistent estimates for the true parameter value $\beta_1$. However, the pairwise likelihood does not. This is due to that fact that the two different parameter for dependence, $\psi_1$ and $\psi_2$, induce some constraints on the true mean parameters $(\beta_0, \beta_1)$. This changes the meaning of the mean parameters that are represented by the pairwise likelihood. %Other composite likelihood approaches, e.g. a composite conditional likelihood may be suitable for this type of problem, however, evaluation of this is beyond the scope of this document. %it is not investigated here

\begin{table}[h!] \small
 \caption{Family study with $k=5$. The maximum profile composite likelihood estimates of $\beta_1$, using independent and pairwise likelihood methods, under a moderate  dependence parameter $\psi=3$ and with different number of families ($n$) and and $\beta_0=-1$, $\beta_1=2$. \label{table:family_k5}}
 \centering
 \begin{tabular}{ r c   ccc    c }
 &&     \multicolumn{3}{c}{$\hat{\beta_1}_{CL_p}$}  &  \\[1ex]   \hline
 n &&  &   independent    &   pairwise    \\[1ex] \hline 
30  &&  &   2.062  & 1.938     \\ 
100  &&  &2.022  & 1.915      \\
300  &&   &2.003  & 1.898     \\ 
500 & &   &2.003   & 1.899    \\ 
1000 & & & 2.002  & 1.899     \\ 
\end{tabular}
\end{table} 

In Table \ref{table:sibling_k5_psi}, we see that as sample size increases, the pairwise likelihood provides consistent estimates for the true parameter value $\psi$. 
\begin{table}[h!] \small
 \caption[The maximum profile composite likelihood estimates of $\psi$ for a sibling study with $k=5$.]{Sibling study with $k=5$. The maximum profile composite likelihood estimates of $\delta=\log(\psi)$, using the pairwise likelihood method and with different number of families ($n$) and $\beta_0=-1$, $\beta_1=2$, $\delta=\log(3)=1.099$. \label{table:sibling_k5_psi}}
 \vspace{0in}
 \centering
 \begin{tabular}{ r c ccccc  }
 $n$ & & 30 & 100 & 300 & 500 & 1000    \\[1ex]  \hline 
$\hat{\delta}_{CL_p}$  &&   1.031  &   1.079 & 1.092 & 1.094 & 1.096 \\ 

\end{tabular}
\end{table} 
%\bigskip
%\clearpage

\end{document}